\documentclass{aastex62}
\usepackage{graphicx} 
\usepackage{threeparttable}
\usepackage{multirow}
\usepackage{amsmath}
\usepackage{bbold}

\received{Month DD, 2019}
\revised{XXX, 2019}
\accepted{XXX, 2019}
\submitjournal{PASP}

\shorttitle{Multiple imputation in astrometric analysis}
\shortauthors{R. Claveria et al.}

\begin{document}

\title{Visual binary stars with partially missing data: Introducing
  multiple imputation in astrometric analysis}


\correspondingauthor{Ruben M. Claveria}
\email{rclaveria@ing.uchile.cl}

\author{Ruben M. Claveria} \affil{Departament of Electrical
  Engineering,\\ Faculty of Physical and Mathematical
  Sciences,\\ Universidad de Chile, Av. Tupper 2007, Santiago, Chile}

\author{Rene A. Mendez}
\affiliation{Department of Astronomy\\
 Faculty of Physical and Mathematical Sciences\\
 Universidad de Chile, Casilla 36-D, Santiago, Chile}

\author{Jorge F. Silva}
\affiliation{Departament of Electrical Engineering,\\
Faculty of Physical and Mathematical Sciences,\\
Universidad de Chile, Av. Tupper 2007, Santiago, Chile}

\author{Marcos E. Orchard}
\affiliation{Departament of Electrical Engineering,\\
Faculty of  Physical and Mathematical Sciences,\\
Universidad de Chile,  Av. Tupper 2007, Santiago, Chile}

\begin{abstract}
Partial measurements of relative position are a relatively common
event during the observation of visual binary stars. However, these
observations are typically discarded when estimating the orbit of a
visual pair. In this article we present a novel framework to
characterize the orbits from a Bayesian standpoint, including partial
observations of relative position as an input for the estimation of
orbital parameters. Our aim is to formally incorporate the information
contained in those partial measurements in a systematic way into the
final inference. In the statistical literature, an imputation is
defined as the replacement of a missing quantity with a plausible
value. To compute posterior distributions of orbital parameters with
partial observations, we propose a technique based on Markov chain
Monte Carlo with multiple imputation. We present the methodology and
test the algorithm with both synthetic and real observations, studying
the effect of incorporating partial measurements in the parameter
estimation. Our results suggest that the inclusion of partial
measurements into the characterization of visual binaries may lead to
a reduction in the uncertainty associated to each orbital element, in
terms of a decrease in dispersion measures (such as the interquartile
range) of the posterior distribution of relevant orbital
parameters. The extent to which the uncertainty decreases after the
incorporation of new data (either complete or partial) depends on how
informative those newly-incorporated measurements are. Quantifying the
information contained in each measurement remains an open issue.

\end{abstract}


\section{Introduction}

Incomplete or missing data is not uncommon in observational astronomy
and, indeed, partial measurements are a relatively common phenomenon
when doing astrometry of visual binary stars. Usually, this phenomenon
does not derive from random data loss (as occurs in communication
systems, for example), but rather from the impossibility of resolving
the relative position between components on an image or sequence of
images, e.g., in speckle interferometry, the maximum angular
resolution for detection of a companion may depend on a number of
factors such as the magnitude difference between the pairs (see, e.g.,
Figure~5 in \citet{Toko2018}). Images of visual pairs, obtained by
means of optical telescopes or interferometric techniques, are
expected to display each component as a sharp point over a fainter
background, thus enabling the viewer to identify their position in the
plane of the sky. However, a number of factors may hinder the process
of resolving a binary star: excessive or insufficient exposure time of
the imaging devices, presence of additional noise due to atmospheric
conditions, varying degrees of brightness of the objects being
observed, angular separations beyond the angular resolution of the
telescope, etc. The latter case is arguably the most common source of
partial information: in the vicinity of the periastron, the angular
separation between the primary and secondary star (denoted by $\rho$
hereafter) reaches its minimum values; if those values fall below the
resolution threshold of the imaging device, then the angular
separation is confined to a range (i.e., $\rho \in (0, \rho_\textrm{max})$)
instead of being reduced to a single value (as occurs with regular,
successfully resolved measurements), and usually no information about
the position angle of the secondary with respect to the primary
(denoted by $\vartheta$ hereafter) can be inferred either. In certain
cases, databases of visual binary observations also report partial
data of the form $\rho = (0, \infty)$, $\vartheta = \vartheta^{\ast}$,
that is, the angular separation is missing but the position angle is
well-defined. Many example of partially missing data (in $\rho$ or
$\vartheta$) can be found by browsing either through the historical
measurements database of the Washington Double Star Catalogue (WDS
hereafter, \citet{mason20012001}) or the Catalog of Interferometric
Measurements of Binary Stars (INT hereafter, \citet{Hartet2001}), both
diligently curated by the Naval Observatory of the United
States\footnote{Latest versions available at
  http://www.usno.navy.mil/USNO/astrometry/optical-IR-prod/wds.}.   

In a broader context, from population censuses to clinical research
surveys, the presence of incomplete or missing data may occur in a
wide range of statistical situations (a more in-depth introduction to be problem can be found in \citet{dempster1977maximum} and \citet{little1987statistical}, Chapter 1).
Since standard inference methods are conceived to be used with complete data sets, 
missing values imply a challenge for data analysts. Several approaches
can be taken to deal with partial or missing data. The most naive strategy 
is to discard data entries with partial measurements (with the information 
contained in them remaining unused), and then apply a standard technique 
on the subset of complete measurements or, in the best of cases, to use the
partial information only in a qualitative assesment of the
solution. However, the validity and efficiency of complete-data based
methods cannot be guaranteed when data are incomplete
\citep{rubin1976inference}. In the particular case of visual binaries,
the incorporation of incomplete data, however, may have the potential
to reduce the uncertainty about orbital parameters. A hint of that can
be appreciated in the fact that some astronomers use partial data in a
somewhat subjective way: incomplete entries are not included as an
input of the estimation routines (whichever they might be:
least-squares methods, Markov chain Monte Carlo, etc.); however,
within a set of orbit proposals (obtained by means of complete data),
the solutions that violate the ranges imposed by partial information
can be rejected (for example, if $\rho$ is known to be in $(0,
\rho_\textrm{max})$ for certain epoch, but the model predicts an angular
separation significantly larger than $\rho_\textrm{max}$). Our conjecture on
this is that partial measurements may be useful in certain
observational conditions, and that the Bayesian approach offers a
proper statistical framework to include this data and improve the
estimation of orbital parameters.

Due to its relevance for the calculation of stellar masses, the
problem of estimating orbital parameters based on observations of
relative positions of visual binary stars has been largely studied in
the astronomical community. From geometry-based procedures such as the
Thiele-Innes-Van den Bos method \citep{thiele1883neue} or
\cite{docobo1985analytic} to modern Bayesian techniques such as those
presented in \cite{ford2005quantifying}, \cite{lucy2014mass}, or
\cite{sahlmann2013astrometric} (to mention a few), orbit fitting has
been addressed from a wide range of approaches. However, the issue of
partial measurements in orbital estimation has been systematically
ignored. This work presents a novel extension of our previous Markov
chain Monte Carlo-based method to estimate orbital parameters
(\citet{Mendezet2017}), that fully incorporates partial data into the
inference using an imputation approach. The cases addressed are those
mentioned in the preceding paragraph, namely, $\rho \in (0,
\rho_\textrm{max}), \vartheta \in (0, 2\pi)$ and $\rho \in (0, \infty)$,
$\vartheta = \vartheta^{\ast}$. Entries where a Cartesian coordinate
is missing\footnote{Astrometric observations of relative position of
binary stars, whether complete or partial, are typically given in
polar coordinates, as in the WDS or INT catalogues. Observations
derived from the photometric analysis of lunar occultations, on the
other hand, are one example of the rarer event of relative position
measurements being expressed in terms of $\Delta x$ and $\Delta y$
(in \citet{evans1986photoelectric}, for example, the authors present
a list of binary stars discovered by this technique.)}, as well as
cases where the feasible zone of the missing observation has an
arbitrary geometry, are theoretical possibilities; however, they are
not addressed in this work, since they do not usually occur in real
situations. To the best of our knowledge, this is the first time that this technique of multiple imputation is applied to the estimation of orbital parameters. For a brief account of the use of this technique in the wider field of astronomy, see \citet{chattopadhyay2014statistical}, Chapter 6.

The structure of our paper is as follows: In Section~\ref{MI_theory}
we introduce the basic concepts of imputation theory and how this can
be implemented through a Markov chain Monte Carlo approach in a very
generic setting. Then, in Section~\ref{MI_method} we bring down to
earth (or rather, to the heavens), the concepts outlined in the
previous section by describing the methodology for our specific
application: computing orbital elements of visual binaries using
partial data. In Sections~\ref{simbasedtests} and \ref{hu177analysis}
we present the results of applying our imputation methodology to a
controlled (simulated) experiment, and to a real test case for the
visual binary HU177 respectively. Finally in Section~\ref{tantan} we
present our main conclusions. Appendices~\ref{appendix0} and \ref{spc}
provide further details regarding the solution of the orbital elements
and the pseudo-codes used, which can be made available in full upon
request to the main author. Appendix \ref{appendix3} contains relevant details on the technical aspects of the Markov chain Monte Carlo (MCMC hereinafter) techniques used in this work: chain convergence, initialization and burn-in period.

\section{Multiple imputation theory and implementation} \label{MI_theory}

\subsection{A premier on multiple imputation} \label{premier}

A number of techniques specifically aimed at addressing the problem of
missing or partial measurements have been proposed. For example,
\cite{dempster1977maximum} use an Expectation-Maximization (EM
hereinafter) algorithm to calculate maximum likelihood estimates from
incomplete data. Another approach is to fill in the blank spaces due
to missing data with plausible values\footnote{The way in which these
  ``plausible values'' can be generated ranges from replacing blank
  spaces with average values, to problem-specific probability models.}
--referred to as imputations--, thus generating a set of observations
on which standard complete-data based methods can be
applied. Regardless of the specific manner in which imputations are
generated, methods that proceed like that can be labeled as ``single
imputation techniques''. Both EM and single imputation techniques have
certain drawbacks: EM is not suitable when the object of interest is
the likelihood or posterior distribution rather than just a maximizer
(and the curvature at that point, at most); on the other hand, single
imputation techniques omit the sources of uncertainty associated to
the imputation process, which may lead to biased results. These
sources of uncertainty are enumerated as follows: One is the
uncertainty associated to the modelling of the joint distribution
of the response variables $\mathcal{Y}$ (observed and unobserved) and
the missingness indicator $R$ (see Equation~(\ref{joint_prob_mod}));
the second is the uncertainty of the imputation model, assuming that
values of the observed data and the model parameters are known
(associated with the term later identified as \emph{conditional
  predictive} distribution, see the first term inside the integral in
Equation~(\ref{ppd})); the third is the uncertainty about the model
parameters (denoted by $\theta$ in what follows) themselves, i.e.,
$p(\theta|\mathcal{Y})$ \citep{zhang2003multiple}. Further details are
explained throughout this section.

Introduced by \citet{rubin}, the approach known as \emph{multiple imputation} is aimed at performing inference from incomplete data
while taking into account the uncertainty sources that single
imputation techniques ignore. Multiple imputation relies on replacing
missing values with not one, but multiple plausible values, thus
generating several complete data sets which differ from each other
only in the imputed values--entries with complete data remain the
same. These data sets are analyzed individually with standard
techniques and the final inference is performed by combining the
individual results (e.g., by calculating an average).

At this point, certain definitions must be introduced in order to
formalize the problem of missing data and the techniques to address
it. Let $\mathcal{Y}$ be an observation matrix, with each row $y_i$
being a single multivariate observation of a certain dimension drawn
from a probability distribution $p(y|\theta)$ governed by the model
parameter vector $\theta$ (in the application to visual binary stars
the dimension of $y_i$ would be two, corresponding to the angular
separation between the components, and the position angle of the
secondary with respect to the primary, whereas the vector $\theta$
will have as its components the seven orbital parameters, see
Section~\ref{ssec:DynamicsBinaryStar}). The components of the
missingness indicator matrix $R$ are defined as:
\begin{equation}
    r_{ij} = \begin{cases}
    0 \text{ if } y_{ij} \text{ is observed},\\
    1 \text{ if } y_{ij} \text{ is missing}.
    \end{cases}
\end{equation}
Defining the probability of $r_{ij} = 0$ given $y_{ij}$ as $p_{ij}$
(conversely, the probability of $r_{ij}=1$ given $y_{ij}$ would be
$1-p_{ij}$), the value of $R$ is subject to a probability distribution
$p(R|\xi, \mathcal{Y})$ which depends on parameters $\xi$ (i.e., $\xi$
encodes the set of rules that define the matrix $R$). Thus, by virtue
of the chain rule of probability, the joint distribution of
$\mathcal{Y}$ and $R$ can be expressed as:
\begin{equation} \label{joint_prob_mod}
p(\mathcal{Y}, R |\theta, \xi) = p(\mathcal{Y}|\theta) \cdot p(R| \xi,
\mathcal{Y}),
\end{equation}
where the term $p(\mathcal{Y}|\theta)$ is the conditional distribution
of the observations given the model parameters. It has been assumed
that $p(\mathcal{Y}|\theta, \xi) = p(\mathcal{Y}|\theta)$ and that
$p(R | \xi, \theta, \mathcal{Y}) = p(R | \xi, \mathcal{Y})$.
Values within $\mathcal{Y}$ can be split
into $\mathcal{Y}_\textrm{mis}$ (values in $\mathcal{Y}$ such that $r_{ij} =
1$) and $\mathcal{Y}_\textrm{obs}$ (values in $\mathcal{Y}$ such that $r_{ij}
= 0$). Since the conclusions about the target parameters $\theta$ must
be based on the joint probability model
(Equation~(\ref{joint_prob_mod})), the manner in which the missingness
depends on $\mathcal{Y}$ must be taken into account when performing
inference. \cite{little1987statistical} identified three missingness
mechanisms:
\begin{itemize}
    \item Missing completely at random (MCAR): $p(R | \xi,
      (\mathcal{Y}_\textrm{mis}, \mathcal{Y}_\textrm{obs})) = p( R | \xi)$. For
      example, in a clinical trial participants would flip a coin to
      decide whether they fill in a depression survey,
    \item Missing at random (MAR): $p(R | \xi, (\mathcal{Y}_\textrm{mis},
      \mathcal{Y}_\textrm{obs})) = p( R | \xi, \mathcal{Y}_\textrm{obs})$, i.e., the
      occurrence of data loss depends only on the observed
      values. Following the example above, male participants could be
      more likely to refuse to complete the depression survey,
      regardless their individual levels of depression (``they tend to
      skip the survey just because they are male''), and,
    \item Missing not at random (MNAR): $p(R | \xi,
      (\mathcal{Y}_\textrm{mis}, \mathcal{Y}_\textrm{obs})) \ne p( R | \xi,
      \mathcal{Y}_\textrm{obs})$, that is, the occurrence of data loss may
      depend on unobserved values. In the clinical trial mentioned
      above, male participants might be more reluctant to complete the
      survey as their level of depression is higher (``they tend to
      skip the survey because they are depressed'').
\end{itemize}
The terms $\theta$ and $\xi$ are said to be \emph{distinct} if their
joint probability can be expressed as a product of independent
marginal probability density functions (PDFs hereinafter)
\citep{rubin1976inference}. If that condition holds, and if either
MCAR or MAR applies, inferences based on the observed-data likelihood
function $L(\theta, \xi | \mathcal{Y}_\textrm{obs}, R)$ will be the same as
those based on $L(\theta|\mathcal{Y}_\textrm{obs})$. In those cases, the
missingness mechanism is said to be \emph{ignorable}. However, the
precision of the inference thus performed is reduced if a large
portion of the information is missing -- this is precisely what
motivates the multiple imputation approach.

The conditional probability of $\mathcal{Y}_\textrm{mis}$ given
$\mathcal{Y}_\textrm{obs}$ can be obtained by integrating over the parameter
space of $\theta$, that is:
\begin{equation} \label{ppd}
p(\mathcal{Y}_\textrm{mis} | \mathcal{Y}_\textrm{obs}) = \int p( \mathcal{Y}_\textrm{mis} |
\mathcal{Y}_\textrm{obs}, \theta) \, p(\theta | \mathcal{Y}_\textrm{obs} ) \, d\theta.
\end{equation}
Following the terminology proposed by \cite{zhang2003multiple}, the
term $p( \mathcal{Y}_\textrm{mis} | \mathcal{Y}_\textrm{obs}, \theta)$ is identified
as the \emph{conditional predictive} distribution of
$\mathcal{Y}_\textrm{mis}$ given $\mathcal{Y}_\textrm{obs}$ and $\theta$. The term
$p(\mathcal{Y}_\textrm{mis} | \mathcal{Y}_\textrm{obs})$ is identified as the
\emph{posterior predictive distribution} of $\mathcal{Y}_\textrm{mis}$ given
$\mathcal{Y}_\textrm{obs}$, and must be understood as the conditional
predictive distribution averaged over the observed-data posterior
  distribution of $\theta$, $p(\theta | \mathcal{Y}_\textrm{obs} )$. Although
  $p(\mathcal{Y}_\textrm{mis} | \mathcal{Y}_\textrm{obs})$ is seldom found in a
  closed form, expressions for both $p( \mathcal{Y}_\textrm{mis} |
  \mathcal{Y}_\textrm{obs}, \theta)$ and $p(\theta | \mathcal{Y}_\textrm{obs} )$ may
  be found in certain situations. On the other hand, $p(
  \mathcal{Y}_\textrm{mis} | \mathcal{Y}_\textrm{obs}, \theta)$ depends on the
  imputation model adopted. Examples of standard imputation models
  existing in the literature are the \emph{predictive model method}
  \citep{little1987statistical} and the \emph{propensity score method}
  \citep{lavori1995multiple}, which are only applicable when data
  follows a monotone missing pattern\footnote{A monotone missing
    pattern means that if certain datum $y_{ij}$ in the data matrix
    $\mathcal{Y}$ is missing, then all subsequent $y_{ik}$ ($k>j$) are
    also missing.}.

Let $Q$ be any quantity of interest to be estimated from the observed
data. Then, the core of the multiple imputation approach is the
estimation of $Q$ by averaging the completed-data posterior, $p( Q |
\mathcal{Y}_\textrm{obs}, \mathcal{Y}_\textrm{mis})$, over the feasible values of
$\mathcal{Y}_\textrm{mis}$ given $\mathcal{Y}_\textrm{obs}$ (which are represented
by the distribution $p(\mathcal{Y}_\textrm{mis} | \mathcal{Y}_\textrm{obs} )$ and
depend, ultimately, on the chosen imputation model):
\begin{equation}
\label{posterior_for_MI}
p(Q | \mathcal{Y}_\textrm{obs}) = \int p( Q | \mathcal{Y}_\textrm{obs},
\mathcal{Y}_\textrm{mis}) \, p(\mathcal{Y}_\textrm{mis} | \mathcal{Y}_\textrm{obs} ) \, d
\mathcal{Y}_\textrm{mis}.
\end{equation}
This integral can be approximated as the discrete average of the
values of $Q$ obtained from a finite, possibly small, number of data
sets filled in with imputations. \cite{zhang2003multiple} presents a
detailed discussion of how the number of imputations affects the
estimation variance.

\subsection{MCMC for multiple imputation} \label{MI_MCMC}
Markov chain Monte Carlo, usually shortened to MCMC, designates a wide
class of sampling techniques that rely on constructing a Markov chain
whose equilibrium distribution is the same as that from which one
desires to sample. The chain is designed to explore the domain of the
target distribution (in this case, orbital elements) in such a way
that it spends most of the time in areas of high probability
\citep{andrieu2003introduction}. Since the implementation of the
algorithm is essentially independent of the target distribution, MCMC
provides a means to efficiently draw samples from distributions with
complex analytic formulae and/or multidimensional
domains. Furthermore, MCMC does not require a complete knowledge of
the target distribution $p(X)$, but just being able to evaluate $p(X)$
up to a normalizing constant ($X$ refers to a generic variable of
interest, in our case $X=\theta$). For a thorough introduction to this
technique, see the textbooks \citep{brooks2011handbook,gelman2013bayesian}.

Introduced in \cite{metropolis1949monte},
\cite{metropolis1953equation} and subsequently generalized to a larger
class of algorithms in \cite{hastings1970monte}\footnote{The sampling
  technique introduced in that work, known as Metropolis-Hastings
  method, is shown in Algorithm~\ref{alg_MH} and constitutes the
  building block on which most modern MCMC and MCMC-inspired
  strategies rely.}, MCMC techniques have been applied to a wide range
of problems, such as statistical mechanics, optimization or --most
relevant to this work-- Bayesian inference
\citep{andrieu2003introduction}. Within the field of astronomy,
Bayesian inference has become somewhat of a standard approach in the
exoplanet research community since the seminal works of
\citet{ford2005quantifying} and \citet{gregory2005bayesianA}, where
the authors address the problem of estimating the orbital parameters
of an exoplanet from a Bayesian perspective and use MCMC to
approximate their posterior distribution. In the last decade, besides
making up an important part of the Kepler Mission data processing
pipeline \citep{rowe2014validation,gautier2012kepler}, recent findings
such as the celebrated discovery of seven temperate terrestrial
planets orbiting the ultra-cool dwarf star TRAPPIST-$1$
\citep{gillon2017seven} relied to some extent on MCMC. Its use in the
characterization of orbits of binary stars, however, has been more
limited, despite the high degree of formal similarity between
exoplanets and multiple stellar sysrtems. Some examples of works
addressing the estimation of orbital parameters of binary stars from a
Bayesian perspective (but only using complete datasets) are
\citet{sahlmann2013astrometric, burgasser2015radio} and
\citet{Mendezet2017}.

When data loss is not governed by a monotone pattern, imputation
models such as the \emph{predictive model method} and the
\emph{propensity score method} cannot be applied. To input the missing
values in cases with arbitrary missing patterns, more advanced
techniques must be adopted. The data augmentation algorithm introduced
in \cite{tanner1987calculation}, which is formally an MCMC method,
stems as a useful tool for those cases.

The data augmentation algorithm is motivated by the representation of
the desired posterior distribution found in
Equation~(\ref{posterior_for_MI}), where the quantity of interest $Q$
is typically a parameter vector $\theta$. The term $p(\theta |
\mathcal{Y}_\textrm{obs})$ depends, in turn, on $p(\mathcal{Y}_\textrm{mis} |
\mathcal{Y}_\textrm{obs})$. \cite{tanner1987calculation} prove, by using a
fixed point-based argument, that under mild conditions the following
scheme converges to $p(\theta|\mathcal{Y}_\textrm{obs})$:
\begin{enumerate}
    \item Generate $\mathcal{Y}_\textrm{mis}^{(i,1)}, \dots,
      \mathcal{Y}_\textrm{mis}^{(i,m)}$ from the current approximation of the
      predictive posterior distribution, $p_i(\mathcal{Y}_\textrm{mis} |
      \mathcal{Y}_\textrm{obs})$. This can be accomplished by, first, drawing
      a sample $\theta_i$ from $p_{i}(\theta | \mathcal{Y}_\textrm{obs})$;
      and, then, sampling $m$ values of $\mathcal{Y}_\textrm{mis}$ from
      $p(\mathcal{Y}_\textrm{mis}|\theta_{i}, \mathcal{Y}_\textrm{obs})$,
    \item Update the approximation of the desired posterior PDF,
      $p_i(\theta | \mathcal{Y}_\textrm{obs})$, by averaging the results
      obtained from the completed data sets:
    \begin{equation} \label{averagingStep}
        p_{i+1}(\theta | \mathcal{Y}_\textrm{obs}) =
        \frac{1}{m}\sum_{j=1}^{m} p(\theta|\mathcal{Y}_\textrm{mis}^{(i,j)},
        \mathcal{Y}_\textrm{obs})
    \end{equation}
    \item If a stopping criterion has been reached, stop. If not,
      return to step $1$.
\end{enumerate}
Steps 1 and 2 are referred to as I-step (Imputation step) and P-step
(Posterior step), respectively, in analogy to the Expectation and
Maximization steps of the EM algorithm. The evaluation of the function
$p(\theta|\mathcal{Y}_\textrm{mis}, \mathcal{Y}_\textrm{obs})$ in step $2$ is
replaced by a sampling procedure in certain situations (e.g., if
$p(\theta|\mathcal{Y}_\textrm{mis}, \mathcal{Y}_\textrm{obs})$ is known up to
constant, if the distribution is not known in a closed form but it is
possible to sample from it, etc.).

\begin{figure}
\caption{Algorithm 1: MCMC pseudo-code for multiple imputation}
\label{alg_MI_MCMC}
\centering
\includegraphics[width=\textwidth]{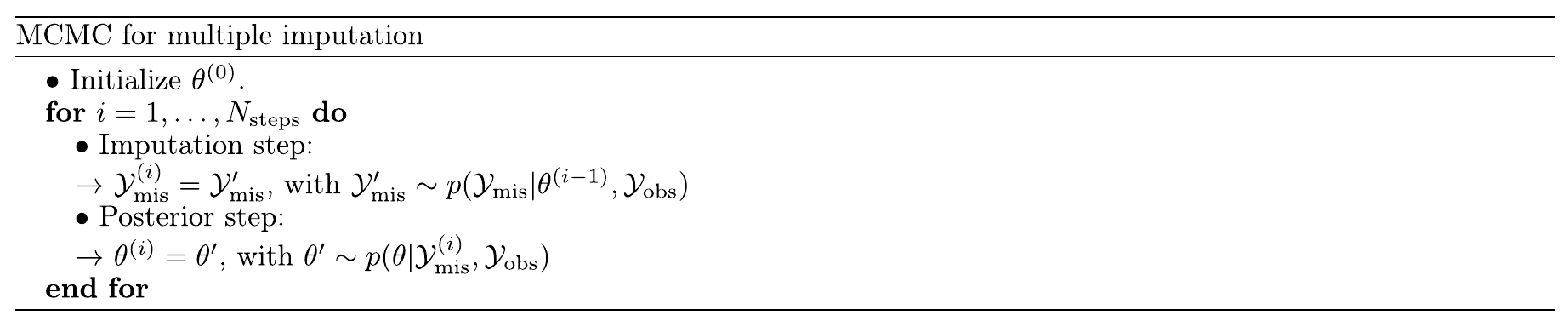}
\end{figure}

\citet{tanner1987calculation} have suggested that even a value as
small as $m = 1$ in Equation~(\ref{averagingStep}) leads to a correct
approximation, in the sense that the average of the posterior
distribution across the iterations --the one obtained in the step 2--
will converge to $p(\theta|\mathcal{Y}_\textrm{obs})$. This motivates the
MCMC scheme actually used in later publications on missing data
\citep{zhang2003multiple} and available on some commercial software
implementations (see, e.g., \citet{yuan2010multiple}). This method is
described in Algorithm \ref{alg_MI_MCMC}: by following this scheme
during a sufficiently large number of iterations, one obtains a
sequence $\{\theta^{(i)}, \mathcal{Y}_\textrm{mis}^{(i)} \}_{i=0,\cdots,
  N_\textrm{steps}}$ whose stationary distribution is $p(\theta,
\mathcal{Y}_\textrm{mis}| \mathcal{Y}_\textrm{obs})$. The chain length, $N_\textrm{steps}$,
is determined empirically based on the desired precision of the
estimated parameters. Marginalizing out $\mathcal{Y}_\textrm{mis}$ yields the
desired posterior distribution $p(\theta|\mathcal{Y}_\textrm{obs})$, whereas
by marginalizing out $\theta$ one obtains the posterior predictive
distribution $p(\mathcal{Y}_\textrm{mis}|\mathcal{Y}_\textrm{obs})$. Note that
$\theta^{(i)}$ is sampled from $p(\theta|\mathcal{Y}_\textrm{mis}^{(i)},
\mathcal{Y}_\textrm{obs})$ instead of $p(\theta|\mathcal{Y}_\textrm{obs})$, since,
given that $m=1$, the averaging step described by
Equation~(\ref{averagingStep}) is not performed.

\section{Methodology as applied to visual binary stars} \label{MI_method}

This section details the implementation of the general technique
described in Section~\ref{MI_theory} applied to the problem of orbital
estimation. Section~\ref{ssec:DynamicsBinaryStar} briefly introduces
the model adopted to mathematically characterize the movement of a
binary star. Section~\ref{ssec:OrbitEst} provides a background on the
use of MCMC for parameter estimation, and details the implementation
used in this work. Finally, Section~\ref{ssec:OrbitEstWithPartialData}
presents the details on how the generic imputation algorithm presented
in Section~\ref{MI_MCMC} is adapted to the specific problem addressed
in this work.

\subsection{Dynamics of a binary star} \label{ssec:DynamicsBinaryStar}

Under the assumption that phenomena such as mass transfer,
relativistic effects or the presence of non-visible additional bodies
do not occur, or have a negligible consequences, we adopt a Keplerian
model to characterize the movement of binary stars. This model
requires seven parameters--the well-known Campbell elements shown in
Equation~(\ref{eqnVIS})--to compute the relative position of the two
components of a binary star for any given epoch $\tau$, namely:
\begin{equation}
    \theta = \{P, T, e, a, \omega, \Omega, i \} \label{eqnVIS},\\
\end{equation}
where $P$, $T$, $e$ correspond to orbital period, time of periastron
passage, and eccentricity respectively; $a$ denotes the semi-major
axis of the elliptical orbit, and the angular parameters $\omega$,
$\Omega$, and $i$ indicate the spatial orientation of the orbit from
the point of the view of the observer. Details about these parameters,
as well as a complete derivation of the Keplerian model, can be
consulted e.g., in \citet{van1967principles}.

Calculating the position $(\rho, \vartheta)$ of the relative orbit (or
the equivalent Cartesian coordinates ($X$, $Y$)) at a certain instant
of time $\tau$ involves a sequence of steps, described in what
follows:
\begin{itemize}
\item Solving Kepler's equation\footnote{We use a Newton-Raphson
  routine to solve this equation numerically.} in order to obtain the
  eccentric anomaly $E$:
\begin{equation}
\label{eqKepler1}
M = 2 \pi (\tau - T)/P = E - e \sin E.
\end{equation}
\item Computing the auxiliary values $x$, $y$, referred to as
  \emph{normalized coordinates} hereinafter:

\begin{eqnarray}
\label{normcoordinates} x(E) &= \cos E - e,\\
\nonumber y(E) &= \sqrt{1-e^2} \sin E.
\end{eqnarray}

\item Determining the Thiele-Innes constants:
\begin{eqnarray}
\label{TI_constants}
A =& ~a (\cos\omega~\cos\Omega - \sin \omega~\sin\Omega~\cos i),\\
\nonumber B =& ~a (\cos\omega~\sin\Omega + \sin \omega~\cos\Omega~\cos i),\\ 
\nonumber F =& ~a (-\sin\omega~\cos\Omega - \cos \omega~\sin\Omega~\cos i),\\
\nonumber G =& ~a (-\sin\omega~\sin\Omega + \cos \omega~\cos\Omega~\cos i).
\end{eqnarray}
\item Calculating the position in the apparent orbit as:
\begin{eqnarray}
\label{proj_TI}
X &= B x + G y\\
Y &= A x + F y. \nonumber
\end{eqnarray}
\end{itemize} 
\noindent By virtue of the Thiele-Innes representation, where the
parameter set $\{P,T,e,a,\omega, \Omega, i \}$ is replaced by the
equivalent $\{P, T, e, A, B, F, G\}$, the dimension of the feature space
of this problem is reduced from 7 to 3: by exploiting the linear
dependency of the Thiele-Innes constants (Eq. \ref{TI_constants}) with
respect to the \emph{normalized coordinates} $x$, $y$ (which in turn
depend on $P$, $T$, $e$ and epochs of observation $\{\tau_i\}_{i = 1,
  \dots, N}$), a least-squares estimate of parameters $A, B, F, G$ can
be obtained from the observations directly by simple matrix
algebra, a procedure that is outlined in Appendix \ref{appendix0}. The
approach described here is well-known within the astronomical
literature, being \citet{hartkopf1989binary, pourbaix1994trial,
  lucy2014mass}, and \citet{Mendezet2017} some examples of its use.

\subsection{Bayesian estimation of orbital parameters} \label{ssec:OrbitEst}

Rather than calculating a point estimate of the orbital elements of a
binary star, the Bayesian approach intends to characterize the degree
of knowledge about those parameters $\theta$, given a set of
observations ($\mathcal{Y}$). This knowledge is represented by a
PDF--the \emph{posterior} parameter distribution
$p(\theta|\mathcal{Y})$-- which, by virtue of the Bayes theorem
(Equation~(\ref{theo_bayes})), is the result of upgrading the previous
knowledge about the variables of interest--represented by the
\emph{prior} distribution $p(\theta)$--with up-to-date measurements,
which modify the current level of knowledge through the likelihood
function $p(\mathcal{Y}|\theta)$. The term $p(\mathcal{Y})$ does not
depend on $\theta$, and thus is dismissed as a mere normalization
constant in sampling applications.
\begin{equation}\label{theo_bayes}
p(\theta|\mathcal{Y}) = \displaystyle \frac{p(\mathcal{Y}|\theta)\cdot
  p(\theta)}{P(\mathcal{Y})}.
\end{equation}
As mentioned in Section~\ref{MI_MCMC}, calculation of posterior
distributions and the integrals associated, such as marginal
distributions or expected values, may involve a number of
difficulties: multidimensionality, complex analytic formulae,
ignorance of the value of a normalization constant,
etc. \citep{andrieu2001sequential}. Thus, the sampling technique
briefly introduced in Section~\ref{MI_MCMC} in the context of partial
measurements --MCMC-- arises as a practical solution to approximate
those quantities. As a necessary background material for
Section~\ref{ssec:OrbitEstWithPartialData}, this section explain how
MCMC-based approximations of posterior PDFs is implemented in the
context of orbital parameters of visual binary stars with complete
measurements:

\begin{itemize}
\item The parameters of interest are the orbital elements contained in
  Equation~(\ref{eqnVIS}). The Thiele-Innes representation is used:
  $\theta = [P, T, e, B, A, F, G]$, with vector $\theta$ separated
  into non-linear and linear elements: $\theta_{NL} = [P, T, e]$
  (those on which MCMC perform its ``exploration'') and $\theta_{L} =
  [B, A, F, G]$ (calculated by least-squares in the manner described
  in Appendix \ref{Appendix1}). Term $\theta_L$ indicates the
  ``linear'' section of the vector of orbital parameters, which can be
  obtained by matrix algebra.
\item According to \cite{ford2005quantifying}, it is common practice
  to choose a non-informative prior that is uniform in the logarithm
  space for positive definite magnitudes (for example, period $P$ and
  semi-major axis $a$), following certain scaling arguments presented
  in \cite{gelman2003bayesian}. In the context of orbital estimation,
  this approach has also been used in, for example,
  \cite{gregory2005bayesianA}, \cite{sahlmann2013astrometric}, and
  \cite{lucy2014mass}. It must be noted, however, that using unbounded uniform priors 
  (in either the original or the logarithmic domain) leads to improper
  distributions. In this work, this problem is circumvented by limiting the domain 
  of the target variable: in sections \ref{hu177analysis}, \ref{simbasedtests}, $P$ values 
      out of the range $[P_\textrm{min}, P_\textrm{max}]~yr$ are rejected\footnote{This means that the prior density of $P$, $p(P) \propto 1/P~\mathbb{1}_{[P_\textrm{min}, P_\textrm{max}]}$ (which is equivalent to using $\log P \sim \mathcal{U}(\log P_\textrm{min}, \log P_\textrm{max})$ on the grounds of change of variables for probability densities), may be integrable and lead to a proper posterior distribution \citep{gelman2006prior}.}. From a practical point of view, the $\log P$ representation
   increases the rate of convergence in systems for
  which the period is not well constrained. In exploration-based
  methods such as MCMC, the use of a Gaussian in the logarithm of $P$
  as a proposal distribution has been proven to be useful
  (\cite{gregory2005bayesianA}), hence this is the approach adopted in
  this work. 
\item The target distribution $f(\cdot)$ of our MCMC routine is the
  posterior distribution of orbital parameters,
  $p(\theta|\mathcal{Y})$, which has the canonical form $prior \times
  likelihood $. Terms from the prior PDF can be dropped, as uniform
  priors were used for $T$, $\log P$ and $e$. Thus, the likelihood
  function can be directly utilized to evaluate the
  Metropolis-Hastings ratio. Assuming individual Gaussian errors for
  each observation, the likelihood function of a set of parameters
  $\theta$ is computed as:

\begin{eqnarray}
\nonumber f(\theta) = p(\theta| \mathcal{Y}) \propto \displaystyle
\exp \biggl(-\frac{1}{2} \Bigl(\sum_{k=1}^{N_{x}}
\frac{1}{\sigma_x^2(k)} [X_\textrm{obs}^{(k)} - X_\textrm{model}(\theta, \tau_k)]^2
+ \\\label{likelihood_function_astro}
\sum_{k=1}^{N_y} \frac{1}{\sigma_y^2(k)} [Y_\textrm{obs}^{(k)} -
  Y_\textrm{model}(\theta,\tau_k)]^2\Bigr)\biggr),
\end{eqnarray}
where $\mathcal{Y}$ is the set of observed values of relative
position ($X_\textrm{obs}$, $Y_\textrm{obs}$). Terms $\sigma_x^2(k), ~\sigma_y^2(k)$ 
indicate the measurement errors at epoch $\tau_k$. Positions $X_\textrm{model}$,
$Y_\textrm{model}$ are calculated according to
$\tau_k$ and the parameter values contained in $\theta$, following the
formulae introduced in Section~\ref{ssec:DynamicsBinaryStar}. Thus,
each value of $\theta$ explored on a run of the MCMC estimation
routine is evaluated according to
Equation~(\ref{likelihood_function_astro}).
\item The specific formulation of MCMC used in this work is the
  well-known Gibbs sampler. First introduced by
  \cite{geman1984stochastic}, the Gibbs sampler relies on sequentially
  sampling each component of the feature space according to their
  conditional distributions (see Algorithm \ref{alg_Gibbs}). On the
  long run, such a scheme is equivalent to drawing samples from the
  joint posterior distribution. Since conditional distributions
  of individual components are not necessarily known in a closed form or easy
  to sample, some of them can be sampled through Metropolis-Hastings steps, as shown in Algorithm \ref{alg_MH_Gibbs}. This scheme, usually referred to as ``Metropolis-Hastings-within-Gibbs \citep{tierney1994markov}, is close in form to the actual implementation used in
  this work. Arguably one of the most well-known MCMC algorithms, and
  forming the basis of the popular software package BUGS
  \citep{spiegelhalter1996bugs}, the Gibbs sampler has been applied to
  a wide variety of problems. Examples of its use in orbit
  characterization can be found in
  \citet{ford2005quantifying,burgasser2015radio}, and
  \citet{Mendezet2017}.
\end{itemize}

\subsection{Orbit estimation with partial measurements}
\label{ssec:OrbitEstWithPartialData}

\begin{figure}
\caption{MCMC with multiple imputation for orbital fitting}
\label{alg_MI_MCMC_2}
\centering
\includegraphics[width=\textwidth]{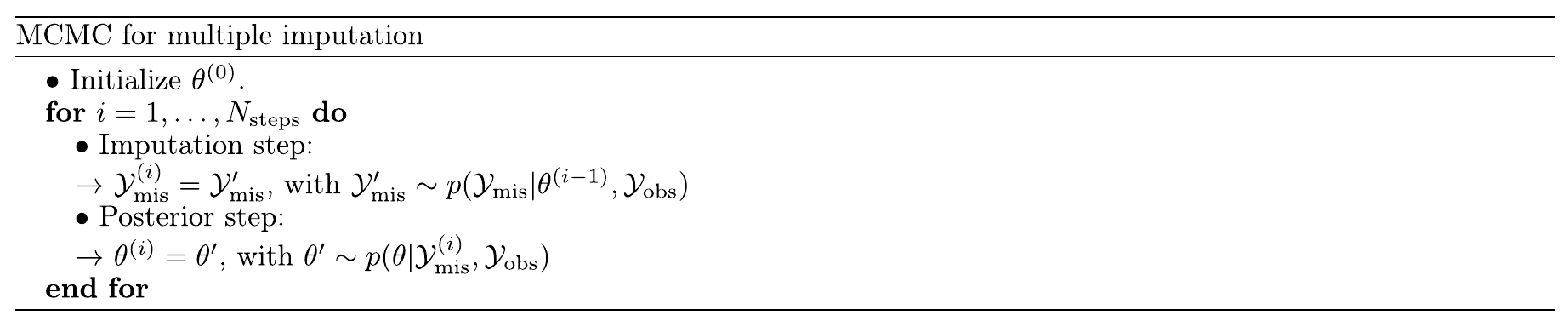}
\end{figure}

A fundamental aspect of imputation-based techniques is determining,
based on reasonable assumptions, how missing values
($\mathcal{Y}_\textrm{mis}$), observed values ($\mathcal{Y}_\textrm{obs}$) and the
target parameters ($\theta$) depend on each other. This involves
defining probability distributions relating each set of values (i.e.,
how $\mathcal{Y}_\textrm{mis}$ depends on $\mathcal{Y}_\textrm{obs}$, how
observations $\mathcal{Y} = \mathcal{Y}_\textrm{obs} \cup \mathcal{Y}_\textrm{mis}$
depend on $\theta$, etc.). The following list details the forms that
the probability distributions involved in the estimation routine
described in Algorithm \ref{alg_MI_MCMC_2} adopt in this particular
problem:
\begin{itemize}
\item Imputations must be drawn from $p(\mathcal{Y}_\textrm{mis} |
  \mathcal{Y}_\textrm{obs})$, but this distribution is rarely sampled
  directly. As explained in Section~\ref{MI_MCMC}, samples from
  $p(\mathcal{Y}_\textrm{mis} | \mathcal{Y}_\textrm{obs})$ can be obtained by a
  two-step procedure that involves drawing values from
  easier-to-sample distributions: first, getting $\theta$ from
  $p(\theta | \mathcal{Y}_\textrm{obs})$; then, conditional to the obtained
  value, generating a sample of $\mathcal{Y}_\textrm{mis}$ from
  $p(\mathcal{Y}_\textrm{mis}|\theta, \mathcal{Y}_\textrm{obs})$. Repeating that
  procedure a large number of times is equivalent to integrating
  $p(\mathcal{Y}_\textrm{mis}|\theta, \mathcal{Y}_\textrm{obs}) \cdot p(\theta |
  \mathcal{Y}_\textrm{obs})$ over $\theta$, which yields $p(\mathcal{Y}_\textrm{mis}
  | \mathcal{Y}_\textrm{obs})$. In the context of orbital fitting, the
  measurements $\mathcal{Y}$ correspond to observations of relative
  position: $\{\tau_i, \rho_i, \vartheta_i\}_{i=1,\dots,N_\textrm{obs}}$. The
  set of measurements $\mathcal{Y}$ can be split into
  $\mathcal{Y}_\textrm{obs}$ (observed values) and $\mathcal{Y}_\textrm{mis}$
  (missing values). Entries in $\mathcal{Y}_\textrm{obs}$ have a well-defined
  value assigned to each field ($\tau$, $\rho$, $\vartheta$, plus
  standard deviation of the observational error, $\sigma_\rho$),
  whereas entries in $\mathcal{Y}_\textrm{mis}$ have one of the forms
  described before: at a given epoch $\tau$, either $\rho \in (0,
  \rho_\textrm{max}), \vartheta \in (0, 2\pi)$ or $\rho \in (0, \infty)$,
  $\vartheta = \vartheta^{\ast}$.
\item The term $p(\theta |\mathcal{Y}_\textrm{mis}, \mathcal{Y}_\textrm{obs})$,
  referred to as \emph{posterior predictive distribution} in the
  context of multiple imputation, is simply the posterior PDF of
  $\theta$ given a complete set of observations, $p(\theta
  |\mathcal{Y})$, as in Equation~(\ref{likelihood_function_astro}).
  However, the use of Equation~(\ref{likelihood_function_astro}) as a
  target distribution in the presence of partial measurements is
  slightly different from its use in settings with complete
  observations, since $\mathcal{Y}$ is generated on each iteration by
  filling partial measurements $\mathcal{Y}_\textrm{mis}$ with samples
  $\mathcal{Y}^{\prime}_\textrm{mis}$ extracted from the conditional
  predictive model $p( \mathcal{Y}_\textrm{mis} | \mathcal{Y}_\textrm{obs}, \theta)$
  instead of being a fixed array of values. Let $N_\textrm{mis}$ be the
  number of partial observations, and let $\tau_\textrm{mis}^{(j)}$ denote an
  epoch where a partial observation occurs, with $j = 1, \dots,
  N_\textrm{mis}$. Then, the imputation corresponding to epoch
  $\tau_\textrm{mis}^{(j)}$ in the $i$-th iteration of the MCMC algorithm is
  denoted $\mathcal{Y}_\textrm{mis}^{(i,j)}$, and has the form
  $\mathcal{Y}_\textrm{mis}^{(i,j)} = (\rho^{(i,j)}, \vartheta^{(i,j)})$
  (polar coordinates) or $\mathcal{Y}_\textrm{mis}^{(i,j)} = (X^{(i,j)},
  Y^{(i,j)})$ (Cartesian coordinates). The term
  $\mathcal{Y}_\textrm{gen}^{(i)}$ denotes the set with complete data
  generated in iteration $i$, that is, the union of successfully
  resolved measurements of relative position (whose value is fixed)
  and the particular values that have been imputed on the $i$-th
  iteration: $\mathcal{Y}_\textrm{gen}^{(i)} = \mathcal{Y}_\textrm{obs} \cup
  \mathcal{Y}_\textrm{mis}^{(i)}$ $=$ $\{\vec{X}_\textrm{obs}, \vec{Y}_\textrm{obs}\} \cup
  \{\vec{X}_\textrm{mis}^{(i)}, \vec{Y}_\textrm{mis}^{(i)}\}$. The specific scheme
  to generate $\mathcal{Y}_\textrm{mis}$ values is explained in the next
  bullet point.
\item The expression $p( \mathcal{Y}_\textrm{mis} | \mathcal{Y}_\textrm{obs},
  \theta)$, identified as \emph{conditional predictive} distribution
  of $\mathcal{Y}_\textrm{mis}$ given $\mathcal{Y}_\textrm{obs}$ and $\theta$, is a
  key term in the imputation scheme, since it is the distribution from
  which the values to fill the missing observations are finally drawn
  (the so-called Imputation step). The localization of a missing
  observation in the plane of the sky depends on four factors:
\begin{itemize}
    \item Geometric restrictions, which are indicated as an input of
      the procedure. As a reminder, this kind of information may take
      the form of either $\rho \in (0, \rho_\textrm{max}), \vartheta \in (0,
      2\pi)$ or $\rho \in (0, \infty)$, $\vartheta =
      \vartheta^{\ast}$. The former involves a circular shape with
      radius $\rho_\textrm{max}$ and the primary as its center; the latter
      corresponds to a line that forms an angle of $\vartheta^{\ast}$
      with the north celestial pole. A clear example of this can be
      seen in Figure~\ref{simulation_MI_orbit},
    \item A given value of orbital parameters $\theta$ (which appears
      explicitly in the expression for the \emph{conditional predictive}
      distribution) that is previously sampled from $p(\theta
      |\mathcal{Y}_\textrm{mis}, \mathcal{Y}_\textrm{obs})$. These orbital
      parameters determine the ``real'' orbit, but the observations
      are also affected by observational noise,
    \item A characterization of the observational error. In this work,
      it is assumed that the difference between the real and the
      observed position follows a Gaussian distribution, therefore it
      can be described by its standard deviation $\sigma_{\rho}$. A
      value for $\sigma_{\rho}$ must be provided for each epoch of
      observation: not only it is fundamental to assign a weight to
      each datum, but also to characterize where missing observations
      may fall, given that the ``real position'' is determined by
      $\theta$, and,
    \item Complete measurements of relative position,
      $\mathcal{Y}_\textrm{obs}$, are indirectly involved, since they
      condition the possibilities of the values that $\theta$ can
      adopt when sampling $p(\theta |\mathcal{Y}_\textrm{mis},
      \mathcal{Y}_\textrm{obs})$ in a previous iteration.  For the sake of
      briefness and clarity, the Imputation step in Algorithm
      \ref{alg_MI_MCMC_2} is not explicitly explained, but shortened
      to an instruction called $generateImputation(\cdot)$. The reason
      for this is that the manner in which imputations are generated,
      intended to replicate the action of drawing samples from the
      distribution $p(\mathcal{Y}_\textrm{mis}|x^{(i-1)},
      \mathcal{Y}_\textrm{obs})$, requires a detailed explanation. The
      procedure is summarized in the points presented next:
\begin{itemize}
    \item If a partial observation is of the type $\rho \in (0,
      \rho_\textrm{max}), \vartheta \in (0, 2\pi)$: first, the predicted
      relative position of the system at epoch $\tau_\textrm{mis}^{(j)}$ is
      calculated according to the orbital parameters contained in
      $\theta^{(i-1)}$; then, an additive Gaussian perturbation with
      standard deviation $\sigma_\rho^{(j)}$ (an approximation of the
      observational error for that particular measurement) is
      applied. If the result meets the geometric constraint of being
      in $(0, \rho_\textrm{max})$, then it is accepted; if not, another
      realization of a Gaussian distribution is performed and added to
      the originally predicted position. This procedure is repeated
      until the geometric restriction is satisfied. In the tests
      performed in this work, the number of repetitions required to
      obtain a value within the radius $\rho_\textrm{max}$ is rarely larger
      than 2.
    \item If a partial observation is of the type $\rho \in (0,
      \infty)$, $\vartheta = \vartheta^{\ast}$: just as in the
      case presented previously, the predicted relative position of
      the system at epoch $\tau_\textrm{mis}^{(j)}$ is calculated according
      to the orbital parameters contained in $\theta^{(i-1)}$ and a
      Gaussian perturbation with standard deviation
      $\sigma_\rho^{(j)}$ is applied on the model-based value. Then,
      instead of evaluating if the obtained value meets the geometric
      restriction --which means, in this case, falling on the line
      defined by $\vartheta = \vartheta^{\ast}$--, that value is
      \emph{projected} on that line. This is equivalent to sample the
      bivariate Gaussian centered at the model-based position with
      $\sigma_X = \sigma_Y = \sigma_\rho^{(j)}$, $\sigma_{XY} = 0$,
      conditional to $\vartheta = \vartheta^{\ast}$
\end{itemize}
\end{itemize}
\end{itemize}

The scheme to sample the extended feature space $(\theta,
\mathcal{Y}_\textrm{mis})$ is summarized in Algorithm \ref{alg_MI_MCMC_2}. As
in the setting with complete measurements, both $Y_\textrm{mis}$ and $\theta$
are sampled by means of the Gibbs sampler. Since the geometric
constraints of the partial observations are not restrictive enough, it
is necessary to have a reasonable knowledge of $\theta$ before running
the algorithm --if the prior guess of $\theta$ is too broad,
imputations will cover a wide zone of the plane of the sky, thus being
not representative of the final distribution of
$\mathcal{Y}_\textrm{mis}$. Because of this, it is recommended to obtain a
preliminary estimation of $\theta$ by running an estimation routine
based on complete measurements only, in order to initialize
$\theta^{(0)}$ in Algorithm \ref{alg_MI_MCMC_2} with a sensible
value.

\section{Simulation-based tests of the inputation scheme} \label{simbasedtests}

This section tests the methodology presented in
Section~\ref{ssec:OrbitEstWithPartialData} on a synthetic data set of
astrometric observations of a binary star. The aim of this experiment
is to assess the effects that the incorporation of partial
measurements would have on parameter estimation, under controlled
conditions. Evaluating the effect of incorporating new measurements
(whether complete or partial) to the problem of orbital fitting has an
intrinsic complexity: it is a well-known fact that not all
observations are equally informative; however, as long as there are no
theoretically-backed tools to quantify the information contained in
each measurement, it is difficult to obtain results that hold for all
cases. For that reason, this study does not intend to be exhaustive
but it aims, rather, at pointing out some of the advantages and
challenges of applying this methodology.

The simulation of artificial observations of relative position are
loosely based on the orbit of the binary star Sirius, which is the
brightest star observed from the Earth. The list presented next
details how the synthetic observations are generated:
\begin{itemize}
    \item Orbital parameters have fixed values: $P = 50.090~[yr]$, $T
      = 1944.220~[yr]$, $e = 0.5923$, $a = 0.750/3^{\prime\prime}$
      (Sirius has a semi-major axis of $a = 7.50^{\prime\prime}$),
      $\omega = 147.27^{\circ}$, $\Omega = 44.57^{\circ}$, $i =
      147.27^{\circ}$. The first epoch of observation, $\tau_1$, is
      set to occur on $1914.00$, then $T^{\prime} = (1944.220 -
      1914.00)/50.090= 30.22/50.090$.
    \item Complete measurements (i.e., those with known scalar values
      for $\rho$ and $\vartheta$) are positioned in such a way that
      they cover the orbit section near the apastron. There are two
      partial observations: one of the type $\rho \in (0, \rho_\textrm{max}),
      \vartheta \in (0, 2\pi)$ and the other of the type $\rho \in (0,
      \infty)$, $\vartheta = \vartheta^{\ast}$; both partial
      measurements are located near periastron. The idea is to mirror
      the fact that, in real settings, the occurrence of partial
      measurements is more probable in zones close to the periastron
      due to the resolution threshold of the imaging devices. The
      values for each measurement are reported in Table
      \ref{partial_table} (epoch, angular separation, position angle
      and observational precision). Synthetic observations were
      generated as a particular realization of a Gaussian noise
      applied on the model-predicted positions (with the parameter
      values indicated previously). Partial measurements were obtained
      by simply dropping one of the components ($\rho$ or
      $\vartheta$).
    \item Observational error follows a Gaussian distribution with
      $\sigma_x = \sigma_y = 0.008^{\prime\prime}$ and $\sigma_x =
      \sigma_y = 0.004^{\prime\prime}$ depending on the observation,
      as indicated in Table \ref{partial_table}.
    \item Parallax is set to $\varpi = 37.9210/3~mas$ (Sirius has a
      parallax of $\varpi = 379.210~mas$). Both $a$ and $\varpi$ were
      modified in order to keep about the same mass sum of Sirius
        ($\sim$3~$M_\odot$), but at the same time admitting more realistic
      values of observational noise. Since Sirius is a bright and
      close star, observations obtained with instruments with $\sigma
      \sim 0.004^{\prime\prime}$ would be excessively small in
      comparison with the apparent size of the orbit. To compensate,
      significantly larger values of $\sigma$ would have had to be
      imposed.
\end{itemize}

\begin{deluxetable}{cccc}
\tablecaption{Ephemerides of a visual binary (synthetic
  data). \label{partial_table}} \tabletypesize{\normalsize}
\tablecolumns{12}
\tablewidth{0pt}
\tablehead{
\colhead{Epoch} & \colhead{$\rho$} & \colhead{$\vartheta$} &
\colhead{$\sigma_\rho$} \\
\colhead{(yr)} & \colhead{$(^{\prime\prime})$} &
\colhead{($^{\circ}$)} & \colhead{$(^{\prime\prime})$}
}   
\startdata
1914.00 & 0.3519 & 10.17 & 0.008\\
1916.50 & 0.3624 & 15.07 & 0.008\\
1919.01 & 0.3748 & 20.69 & 0.008\\
1921.51 & 0.3716 & 24.45 & 0.008\\
1924.02 & 0.3657 & 29.69 & 0.008\\
1926.52 & 0.3651 & 33.57 & 0.008\\
1929.03 & 0.3617 & 39.13 & 0.008\\
1931.53 & 0.3270 & 44.68 & 0.008\\
1944.00\tablenotemark{a} & 0.0976\tablenotemark{b} $~$or $(0, 0.0976)$\tablenotemark{c} & 195.42\tablenotemark{b} $~$or $(0, 360)$\tablenotemark{c} & 0.004\\
1945.00\tablenotemark{a} & 0.1047\tablenotemark{b}$~$ or $(0, \infty)$\tablenotemark{c} & 217.14 & 0.004\\
1956.58 & 0.2460 & 346.73 & 0.004\\
1959.08 & 0.2805 & 356.38 & 0.004\\
\enddata
\tablenotetext{a}{Partial observation.}
\tablenotetext{b}{Used in complete information scenario.}
\tablenotetext{c}{Used in partial information scenario.}
\end{deluxetable}

The experiment consists of three scenarios where the algorithms
proposed in this work are applied. In the first case, partial
observations are discarded altogether, and orbital parameters are
estimated by applying the Gibbs sampler\footnote{This Gibbs sampler
  implemented in this case operates like the Posterior step of
  Algorithm \ref{alg_MI_MCMC_2}.} to the sub-set of complete
observations. In the second setting, the two partial
observations are incorporated by means of Algorithm
\ref{alg_MI_MCMC_2}, thus sampling orbital parameters $\theta$ and
``missing'' observations $\mathcal{Y}_\textrm{mis}$ simultaneously. Finally,
in the third scenario, it is assumed that the partial measurements of
the second scenario are completely known (see Table
\ref{partial_table}). This last setting has the role of a
ground-truth, in the sense that it yields the ``best estimation
possible'' given all the data points (it answers the question of how
the estimation would improve if certain values of the observations
associated to epochs $1944.00$ and $1945.00$ had not been dropped). As
in the first case, estimation is carried out by means of the Gibbs
sampler. The MCMC algorithms were run with the following parameters:
$N_\textrm{steps} = 4\cdot 10^6$ with a thinning factor of $10$ (in order to obtain nearly independent samples for inference); Gaussian proposal distributions $q(x^\prime|x)$
(see Appendix~\ref{spc}) with parameters $\sigma_{\log P} = 0.4$,
$\sigma_{T^{\prime}} = 0.01$, $\sigma_e = 0.01$. A burn-in
period\footnote{Burn-in period is the number of samples discarded at
  the beginning of each chain to ensure that the chain starts from an
  equilibrium point, and it is determined empirically.} of $10^5$
samples was determined based on visual inspection of the parameter
values over the successive iterations of the algorithm (details on the initial state of the Markov chains, burn-in period and convergence diagnostics, as well as the specific priors used in this experiment, are presented in Appendix \ref{appendix3}). In the case of
MCMC with multiple imputation, a Gaussian proposal distribution with
$\sigma_\rho = 0.004^{\prime\prime}$ (the same value ``reported'' as
observational error in Table \ref{partial_table}) was used to generate
samples of $\mathcal{Y}_\textrm{mis}$ in the Imputation step.

\begin{figure*}
\begin{minipage}[!ht]{0.5\linewidth}
\centering
\includegraphics[height=.245\textheight]{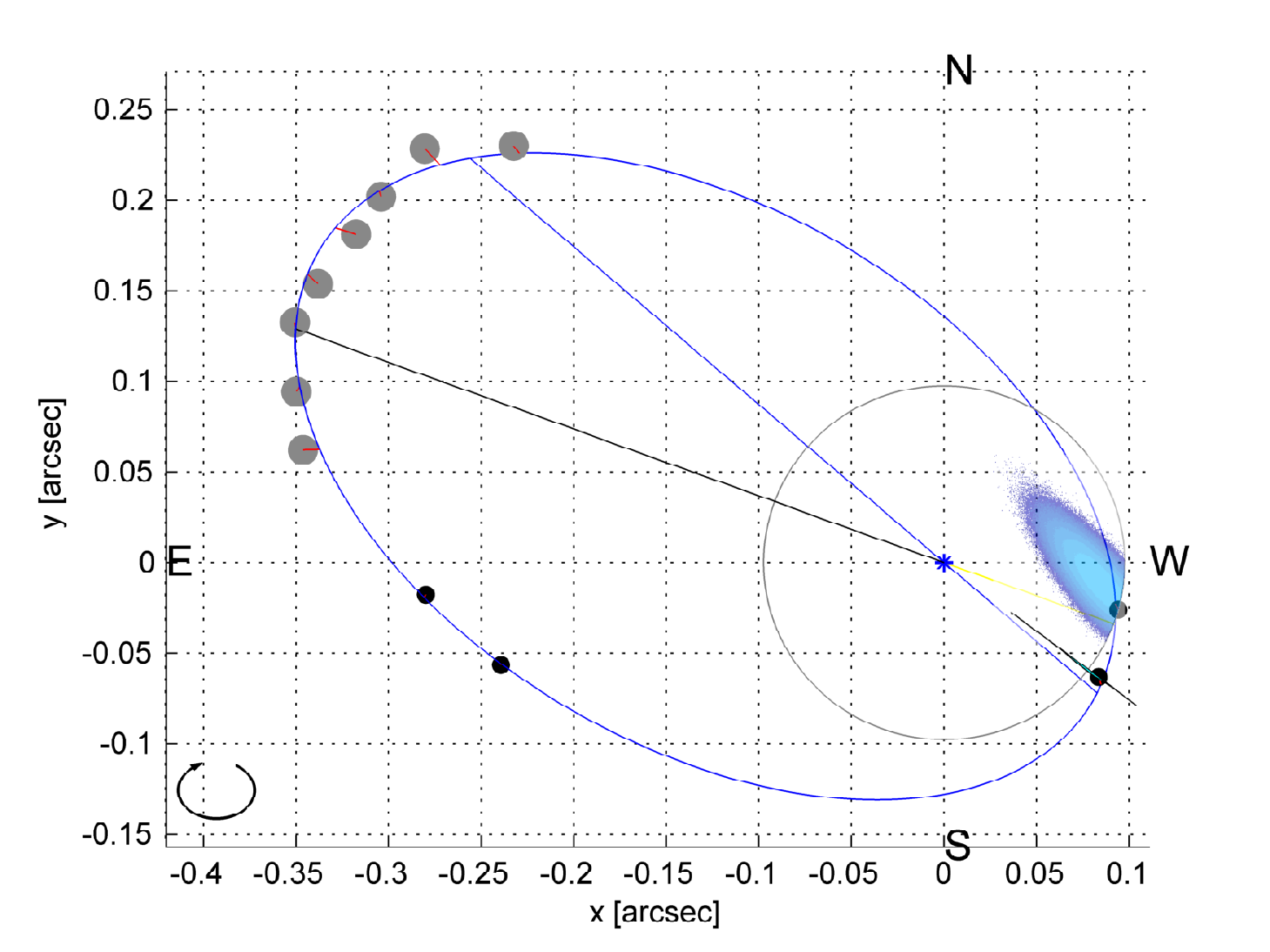}
\end{minipage}
\begin{minipage}[!ht]{0.5\linewidth}
\centering
\includegraphics[height=.245\textheight]{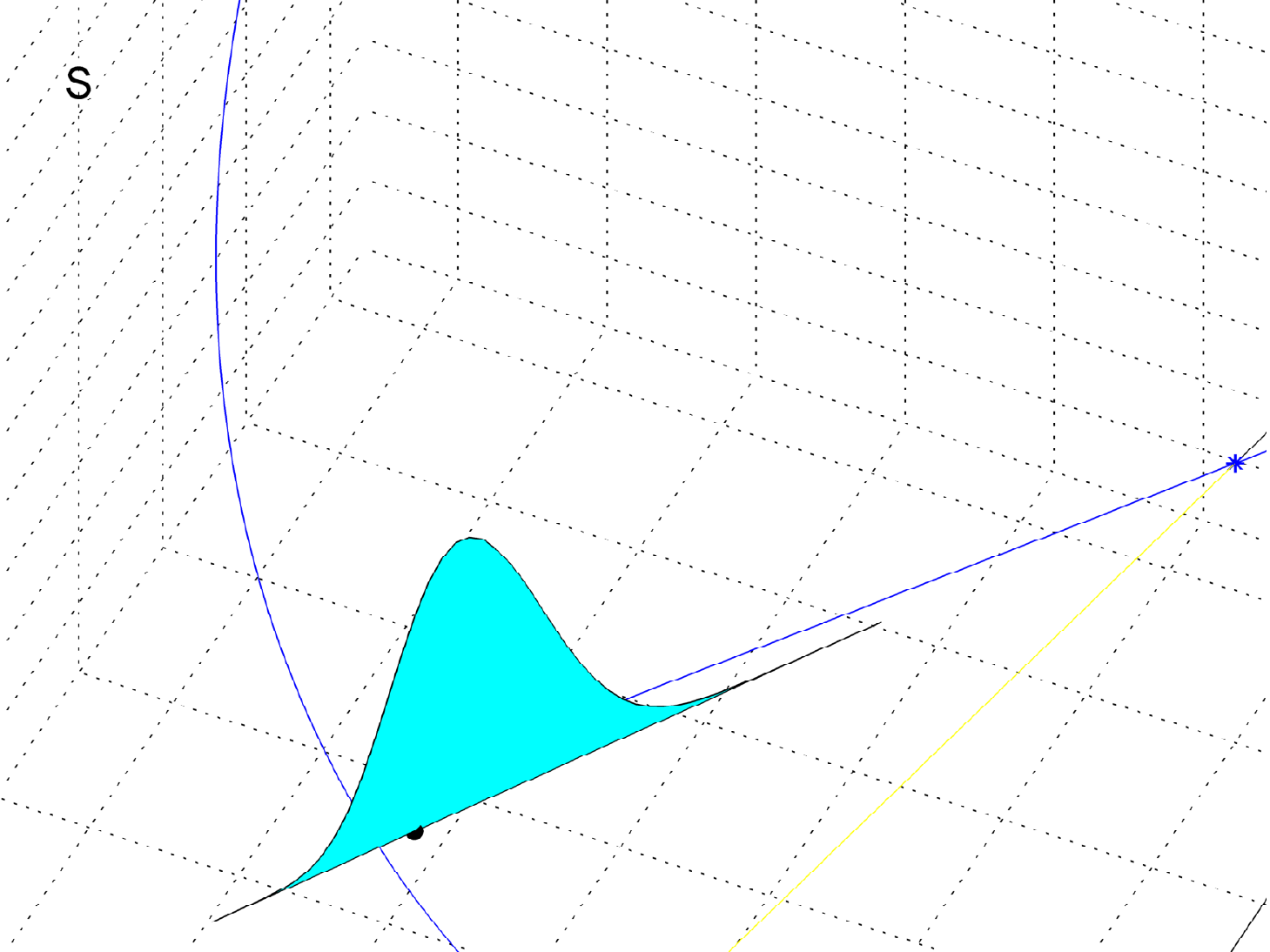}
\end{minipage} 
\caption
{Orbit estimate in presence of incomplete information from our
  synthetic data on Table~\ref{partial_table}. The left panel shows
  the original measurements (grey dots lower weight, black dots higher
  weight, the two black dots near periastron are only included for
  reference purposes, but are not part of the solution with imputed
  values), the maximum likelihood estimate of the orbit (blue line) incorporating the
  incomplete data at epochs 1944.0 and 1945.0, and the zones were
  missing observations may fall (calculated with the proposed
  imputation method). The right panel is a view in detail of the PDF
  for the angular separation $\rho$ along the line $\vartheta =
  217.14^{\circ}$ for the case $\rho \in (0, \infty)$ at epoch 1945.0
  \label{simulation_MI_orbit}}
\end{figure*}

Figure~\ref{simulation_MI_orbit} shows the synthetic measurements
  and the resultant estimated orbit (a maximum likelihood estimate is
  used) including the partial measurements through our imputation
  scheme (i.e., our second scenario described above). PDFs of partial
measurements $\mathcal{Y}_\textrm{mis}$ (obtained in the second scenario) are
superimposed on the graph that displays the complete observations and
the estimated orbit. Those PDFs indicate where the partial
observations may fall on the plane of the sky, given the available
data. Depending on the type of partial measurement, the domain of
these PDFs can take the form of an area bounded by a circle
(case $\rho \in (0, \rho_\textrm{max})$; on
Figure~\ref{simulation_MI_orbit}, lighter shades of blue indicate
higher values of the PDF) or a line (case $\vartheta =\vartheta^{\ast}$, see details on the right panel on
Figure~\ref{simulation_MI_orbit}). The original (complete) values of
the partial measurements are displayed as black dots for referential
purposes, but are not actually used in the estimation procedure of the
second scenario.

\subsection{Analysis of results}

\begin{figure*}
\begin{minipage}[!ht]{0.5\linewidth}
\centering
\includegraphics[height=.28\textheight]{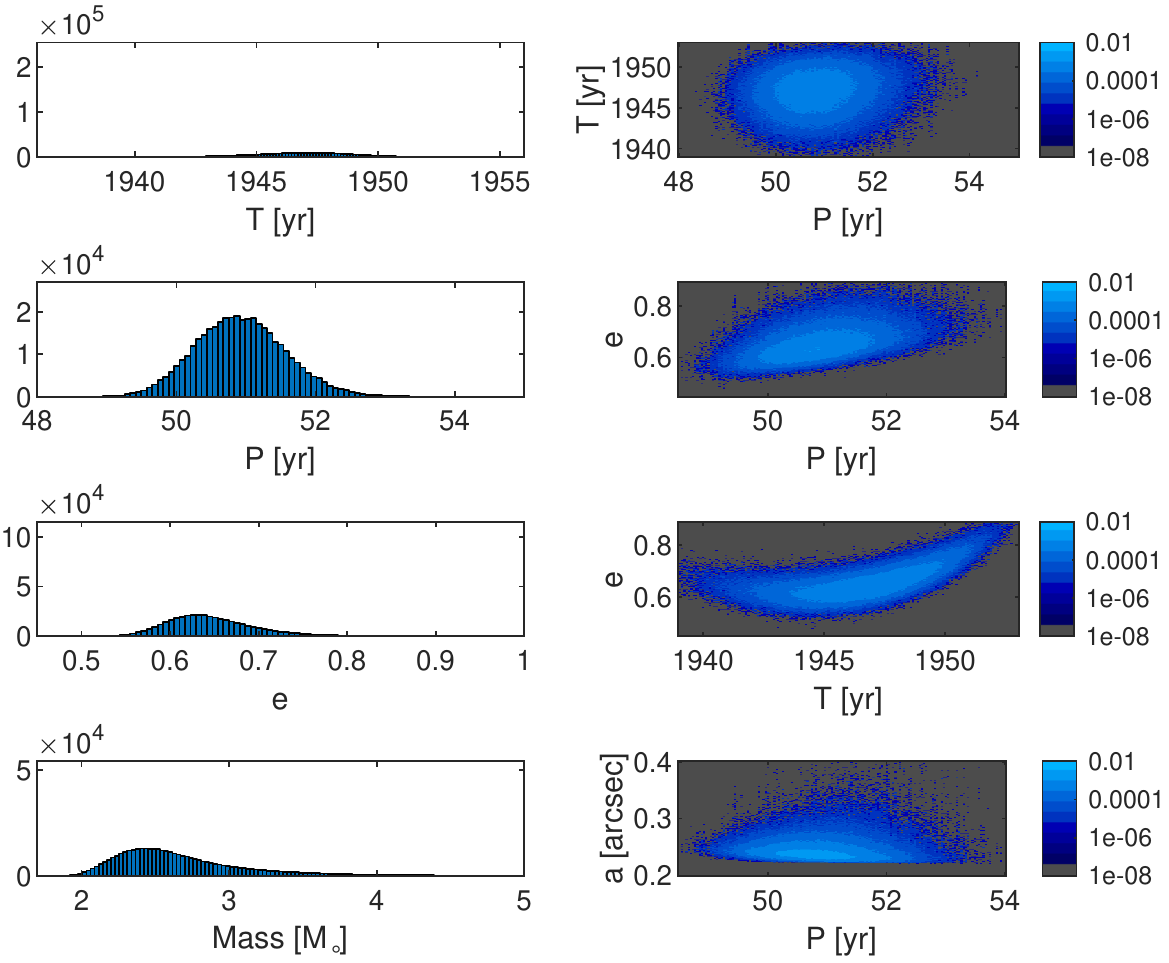}\\
\includegraphics[height=.28\textheight]{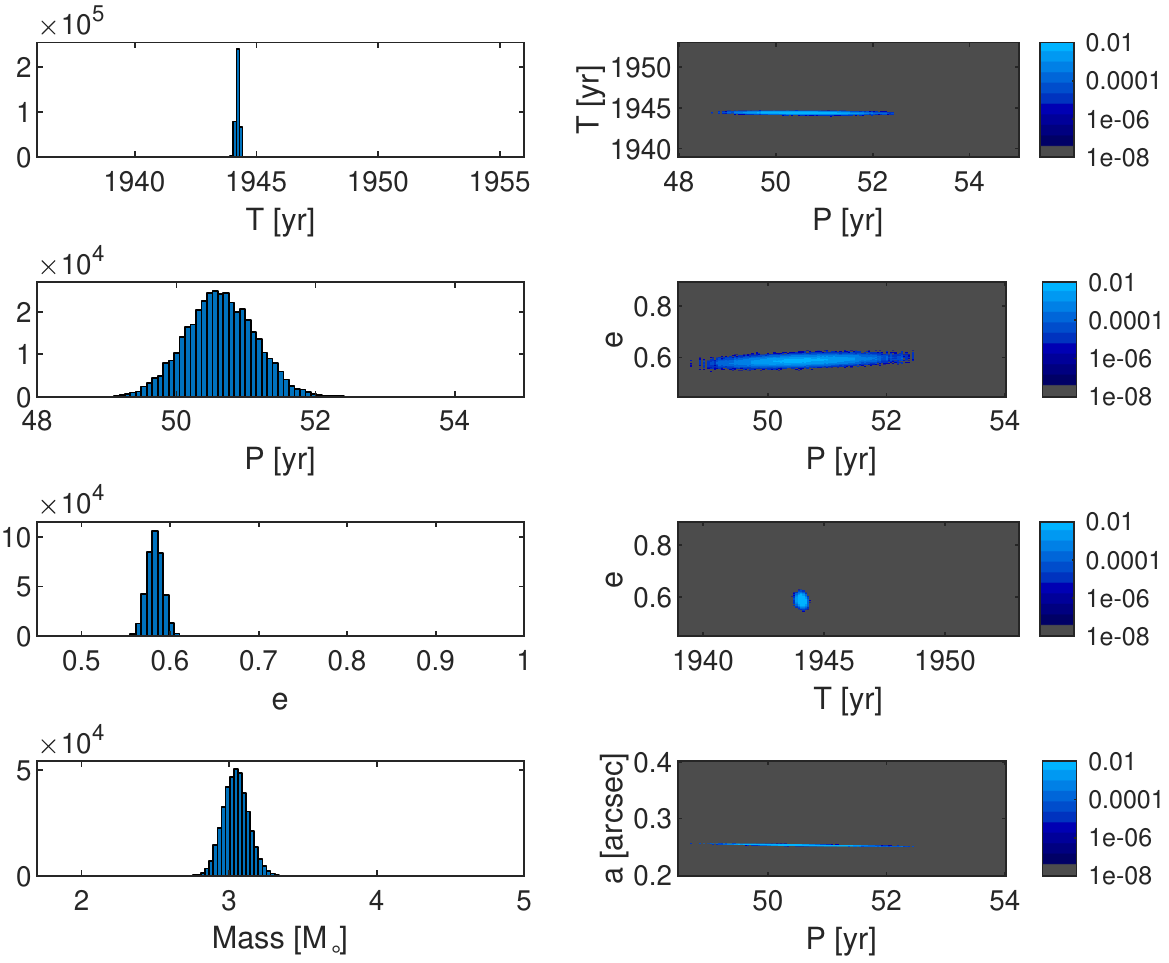}
\end{minipage}
\begin{minipage}[!ht]{0.5\linewidth}
\centering
\includegraphics[height=.28\textheight]{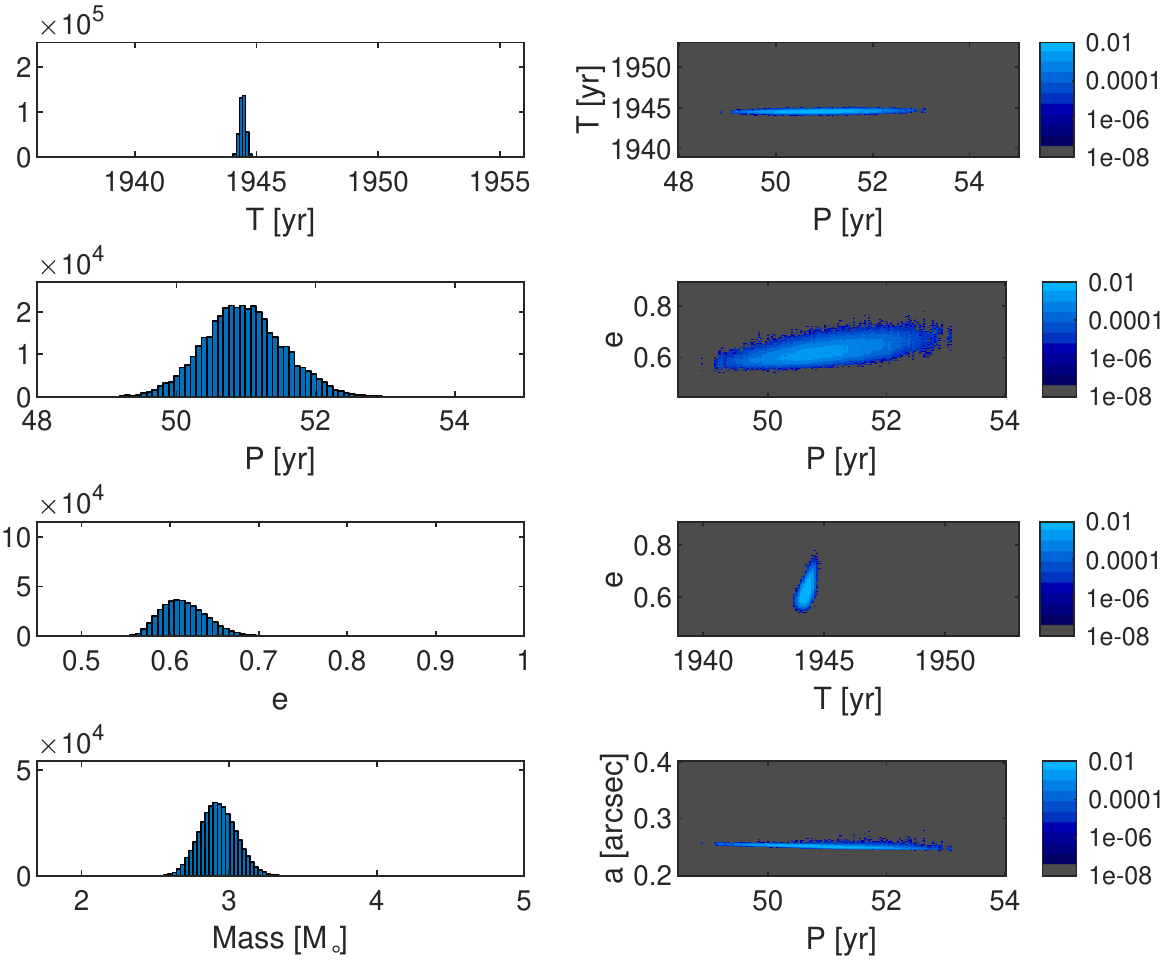}\\
\includegraphics[height=.28\textheight]{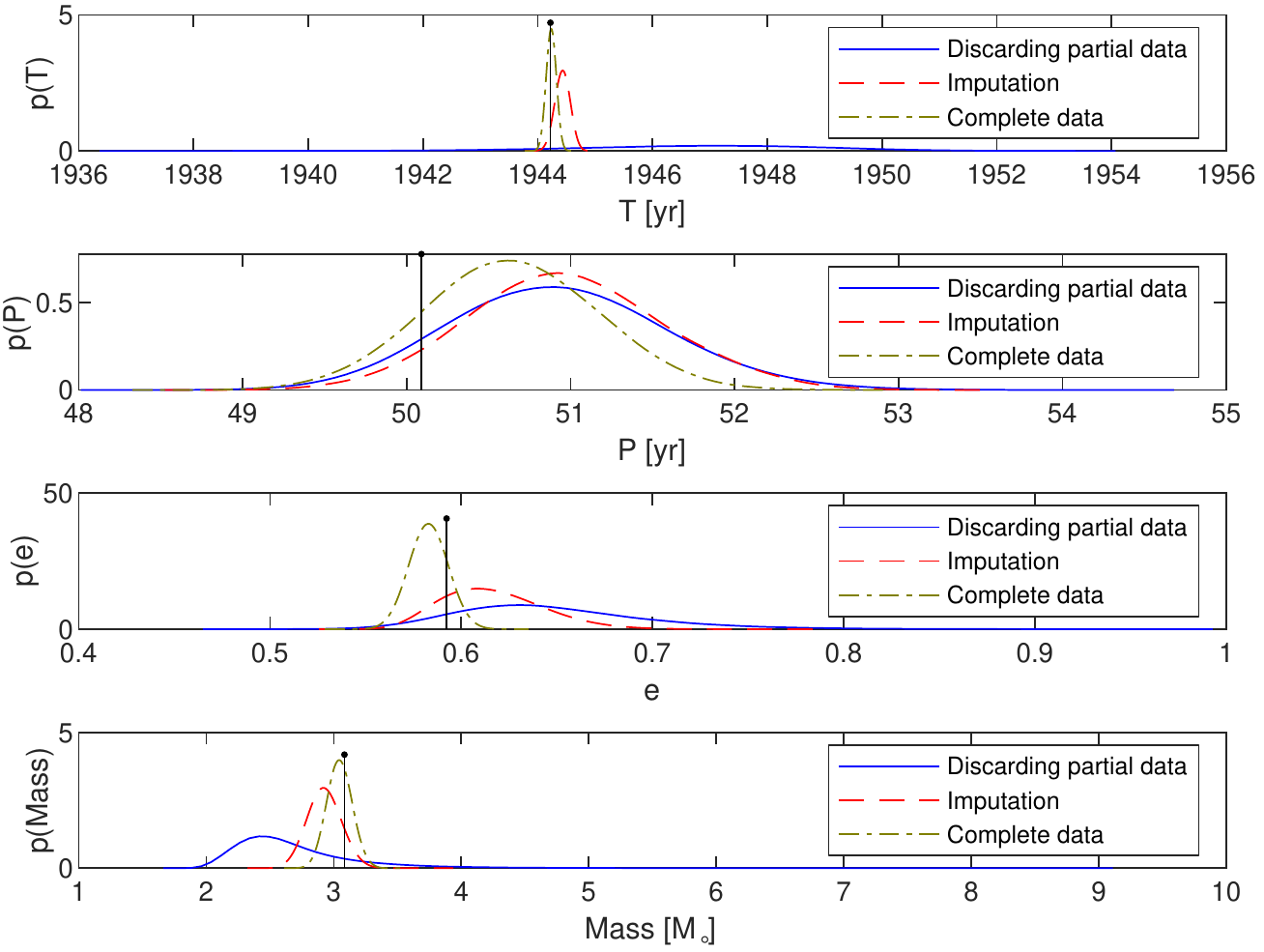}
\end{minipage} 
\caption{Posterior PDFs of orbital parameters from our synthetic data
  on Table~\ref{partial_table}. The four upper left rows show the
  results obtained when completely discarding the partial observations
  at epochs 1944.0 and 1945.0 (i.e., the expected result from a
  standard analysis); the four upper right rows show the results
  obtained when partial information has been incorporated using our
  imputation methodology; the four lower left rows show the results
  obtained using complete information (i.e., as if observations at
  epochs 1944.0 and 1945.0 had measurements for both $\rho$ and
  $\vartheta$ as indicated in Table~\ref{partial_table}, i.e., the
  best possible case from the available data set). Finally, the four
  lower right panels compare the marginal PDFs of the three cases on
  the same figure (the true parameter value is indicated with a black
  vertical bar).\label{synth_test_panels}}
\end{figure*}

Figure~\ref{synth_test_panels} presents the marginal PDFs of the main
orbital parameters ($P$, $T$, $e$, $a$ and total mass) obtained on
each setting, and adds a comparison of the three scenarios on the same
graph. In order to facilitate a visual comparison, histograms are
replaced by kernel-based densities in the lower right panel, although
both convey essentially the same information.

In a certain sense, the general technique described in Algorithm
\ref{alg_MI_MCMC} (being Algorithm \ref{alg_MI_MCMC_2} an
implementation for the specific problem of visual binaries) is based
on augmenting the parameter vector with ``slots'' for the missing
measurements, thus estimating both of them simultaneously. The
simplest, most evident conclusion of the experiment is that the scheme
developed in this paper achieves the objective of sampling both
quantities of interest in parallel (i.e., $\theta$ and
$\mathcal{Y}_\textrm{mis}$), regardless of the quality of the results from an
estimation point of view --how the incorporation of additional data
affects the estimation is a rather different
question. Figure~\ref{simulation_MI_orbit} gives a demonstration of
the capability of this method to characterize the uncertainty about
the missing observations; Figure~\ref{synth_test_panels}, on the other
hand, reveals how the resulting posterior PDFs of orbital parameters
differ from each other on each scenario.

As can be seen in Figure~\ref{simulation_MI_orbit}, this ``multiple imputation via MCMC'' scheme manages to characterize the feasible
values of the partial measurements on the plane of the sky,
simultaneously integrating the geometric restrictions of partial
measurements and the information about $\theta$ conveyed by complete
measurements. It is worth noting that, as values of
$\mathcal{Y}_\textrm{mis}$ and $\theta$ influence each other along the
succession of Imputation and Posterior steps, the final results for
both $\mathcal{Y}_{mis}$ and $\theta$ are different from those that
would have arisen from each source of knowledge on its own: according
to geometric restrictions, feasible values would be evenly distributed
within either a circular zone ($\rho \in (0, \rho_\textrm{max})$, see left
panel of Figure~\ref{simulation_MI_orbit}) or an infinite line
($\vartheta = \vartheta^{\ast}$, see right panel of
Figure~\ref{simulation_MI_orbit}); according to complete measurements,
the location of partial observations would be defined by an \emph{a
  priori} set of feasible orbits (which induce ``real'' positions in
the epochs of interest) plus some kind of observational noise. That
scheme, however, ignores the influence that imputed observations would
exert on the orbital parameters being estimated (in the case of the
$\rho \in (0, \rho_\textrm{max})$ observation, for example, the support of
the resulting PDF covers a much larger area within the circle if
compared with that obtained by means of Algorithm \ref{alg_MI_MCMC_2}
and displayed in the left panel of Figure~\ref{simulation_MI_orbit}).

In Figure~\ref{synth_test_panels}, the posterior PDFs of the target
parameters obtained for each of the scenarios previously described are
displayed for comparison purposes. In general, parameter uncertainty
decreases as more information is incorporated: of course, the tightest
PDFs are obtained in the third scenario (ideal setting with complete
information), whereas the widest, least concentrated PDFs are obtained
when partial information is discarded (first scenario). Results of the
second setting (partial information incorporated via Algorithm
\ref{alg_MI_MCMC_2}) lie somewhere in the middle. The lower right
panel of Figure~\ref{synth_test_panels} is the most explicit graph in
this regard, clearly showing the peaks of the PDFs and how
concentrated they are. However, not all parameters are equally
affected by the incorporation of information (or lack of): while the
uncertainties of both $T$ and (very relevant from the astrophysical
point of view) the mass sum decrease dramatically with the
incorporation of partial data, the marginal PDF of $e$ obtained in the
second scenario resembles more that obtained in the first setting than
that obtained with complete information; the PDFs obtained for $P$, on
the other hand, are quite similar in the three cases. Joint marginal
PDFs undergo a similar change: the distribution of $e$ vs. $T$, for
example, shows a bow-shaped profile in the first scenario, turning
into a convex shape\footnote{In the sense of defining a non-concave
  contour.}  when partial observations are included, and reaching a
Gaussian-like form in the complete information scenario.

In summary, the tests performed in this section, aside from presenting
a demonstration that the methodology developed throughout actually
works as claimed, strongly suggests that the incorporation of partial
information into the analysis tools has the clear potential of
decreasing the orbital parameter estimation uncertainties.

\section{A case study: the visual binary star HU177} \label{hu177analysis}

In this section, the methodology proposed in Section~\ref{MI_method},
and tested in Section~\ref{simbasedtests} with synthetic data, is
applied to a real object: the visual binary star HU177 (WDS
J17305-1446AB = HIP 85679 = HD 158561). This object has been
recently studied (using only complete data) by \citet{Mendezet2017}.
The reason for choosing HU177 for a case study is that, among objects
with partial data, this is one for which the incorporation of partial
measurements may make a relevant difference in terms of the results of
the inference\footnote{In their Catalog of Orbits of Visual Binary
  Stars (available at
  https://www.usno.navy.mil/USNO/astrometry/optical-IR-prod/wds/orb6),
  the US Naval Observatory has devised an orbit ``grading'' system,
  where grade 5 means an indeterminate or very uncertain orbit due to
  poor orbit coverage and/or bad data quality, while grade 1 means a
  definitive orbit. In the WDS Orbit catalogue HU177 has grade 5, but
  \citet{Mendezet2017} proposed promoting it to grade 3 (``reliable'')
  after incorporating more recent interferometric measurements.}. For
objects with a well determined orbit, like STF 2729AB (WDS
J20514-0538AB = HIP 102945 = HD 198571, grade 2,
``good'')\footnote{\citet{Mendezet2017} and the Catalog of Orbits of
  Visual Binary Stars can be consulted for orbit estimates of this
  star.}, for which a large amount of incomplete observations are also
available, the incorporation of partial measurements into the analysis
would hardly affect the value of the estimated parameters and the
uncertainty associated (we have actually verified this by running our
imputation code on this data-set). On the other hand, objects with a
highly indeterminate orbit (grade 5) are equally inadequate for
testing the incorporation of partial measurements, since such a level
of uncertainty would induce spatially disperse imputations (compare
left and right panels of Figure~\ref{comparisonHU177}: with less
information, the area covered by the support of the PDF of the partial
measurement is larger). HU177 has relatively few complete observations
(16 in total), distributed unevenly from 1900 to 2015 and showing
varying degrees of precision. According to the results obtained when
all the 16 complete observations are used to characterize the orbit
(third row in Table \ref{HU177_table_2}), the orbital parameters of
this star are neither tightly constrained nor extremely
indeterminate--this is precisely the scenario in which partial
measurements may make a contribution. The measurements available for
HU177 are given in Table~\ref{TabHU177}, and were kindly provided to
us by Dr. William Hartkopf from the US Naval Observatory as part of
the WDS effort (it includes 15 complete observations, plus one partial
datum at epoch 1991.25 reported by the Hipparcos satellite),
supplemented with our own (complete) measurement at epoch 2015.5409
from the SOAR/HRCam Speckle camera reported in \citet{Tokoet2016}, and
used by \citet{Mendezet2017} to compute an orbit of HU177 (excluding
the partial datum).\newline

\begin{deluxetable}{cccc}
\tablecaption{Observations for the visual binary HU177. \label{TabHU177}}
\tabletypesize{\normalsize}
\tablecolumns{12}
\tablewidth{0pt}
\tablehead{
\colhead{Epoch} & \colhead{$\rho$} & \colhead{$\vartheta$} &
\colhead{$\sigma_\rho$} \\
\colhead{(yr)} & \colhead{$(^{\prime\prime})$} &
\colhead{($^{\circ}$)} & \colhead{$(^{\prime\prime})$}
}
\startdata
1900.54 &  0.370 & 85.2 &  0.250\\
1923.80 &  0.360 & 83.9 &  0.050\\
1936.61 &  0.320 & 67.3 &  0.050\\
1938.30 &  0.340 & 64.8 &  0.050\\
1944.36 &  0.300 & 57.2 &  0.050\\
1957.62 &  0.280 & 33.1 &  0.050\\
1958.45 &  0.220 & 34.2 &  0.050\\
1958.52 &  0.260 & 39.8 &   0.050\\
1962.59 &  0.210 & 28.9 &   0.050\\
1989.3121\tablenotemark{a} &  0.142 & 275.7 &  0.005\\
1991.250\tablenotemark{b} &  $(0,~0.1)$ & $(0,~360)$ & 0.015\\
2007.6013 &  0.205  & 199.8 &  0.015\\
2008.5397 &  0.2217 & 188.5 &  0.001\\
2008.5397 &  0.2234 & 188.2 &  0.003\\
2008.6052 &  0.223  & 188.5 &  0.001\\
2010.5908 &  0.228  & 184.9 &  0.002\\
2015.5409 &  0.2587 & 176.6 &  0.002\\
\enddata
\tablenotetext{a}{Excluded in the HU177$^\ast$ scenario}
\tablenotetext{b}{Partial observation}
\end{deluxetable}

The test performed in this section comprehends two different data
sets: the first is identified by HU177 and contains all the
complete measurements available, plus the partial observation; the
second one, identified as HU177$^{\ast}$, contains all the complete
measurements except for the one near the periastron (epoch
$1989.312$), plus the partial observation (the discarded observation
can be noticed in Figure~\ref{comparisonHU177}). The ``assembly'' of
two separate data sets is carried out in order to better assess the
impact of the partial observation: in presence of a complete
observation occurring closely in time and space (as in data set
HU177), the partial observation may be redundant or even a source of
additional uncertainty (recall that Section~\ref{MI_theory} identifies
three sources of uncertainty associated to the imputation
process). For each data set, two cases are addressed: without and with
multiple imputation. In the first case the partial measurement is
omitted completely, then the data set is analyzed by means of the
Gibbs sampler used in the previous section (the first and third
scenarios described there). The second case is analyzed by means of
Algorithm~\ref{alg_MI_MCMC_2}. Chains were run with the following parameters: $N_\textrm{steps} = 10^7$ (with a thinning factor of $10$ samples); $\sigma_{\log P}$,
$\sigma_{T^{\prime}}$, and $\sigma_{e}$ set to the same values used in
Section~\ref{simbasedtests}. As done for Section~\ref{simbasedtests}, details on prior distributions, chain initialization, burn-in period, convergence diagnostics prior for this experiment are shown in Appendix~\ref{appendix3}. When applying the multiple imputation
technique, the proposal distribution used to generate samples of
$\mathcal{Y}_{mis}$ is set to $\sigma_{\rho} = 0.015^{\prime\prime}$,
based on the observational error associated to the imaging device that
produced the partial measurement. The criteria to fix that value, as
well as the weight assigned to the imputed observations in the
likelihood function, are topics of further research, and will not be
addressed here.

Figure~\ref{comparisonHU177} displays the orbits associated to the
maximum likelihood estimates obtained for data sets HU177$^{\ast}$ and
HU177 with the incorporation of the partial data. The complete
measurements of relative position, as well as the PDF of the partial
observation in each scenario, are shown in the same plane. The maximum
likelihood parameter values corresponding to each of these solutions
are reported in Table~\ref{HU177_table_1}. On the other hand,
Table~\ref{HU177_table_2} reports numerically the results obtained for
the four cases analyzed in terms of their quartiles: each row has two
lines, the first showing the quartiles of the orbital parameters, and
the second line reporting the interquartile range (i.e., the
difference between the 75\% and 25\% percentiles). The interquartile
range is used as a means to assess how the uncertainty on the
corresponding orbital parameter changes when partial information is
incorporated.

\subsection{Results and analysis}

\begin{figure*}
\begin{minipage}[!ht]{0.5\linewidth}
\centering
\includegraphics[height=.25\textheight]{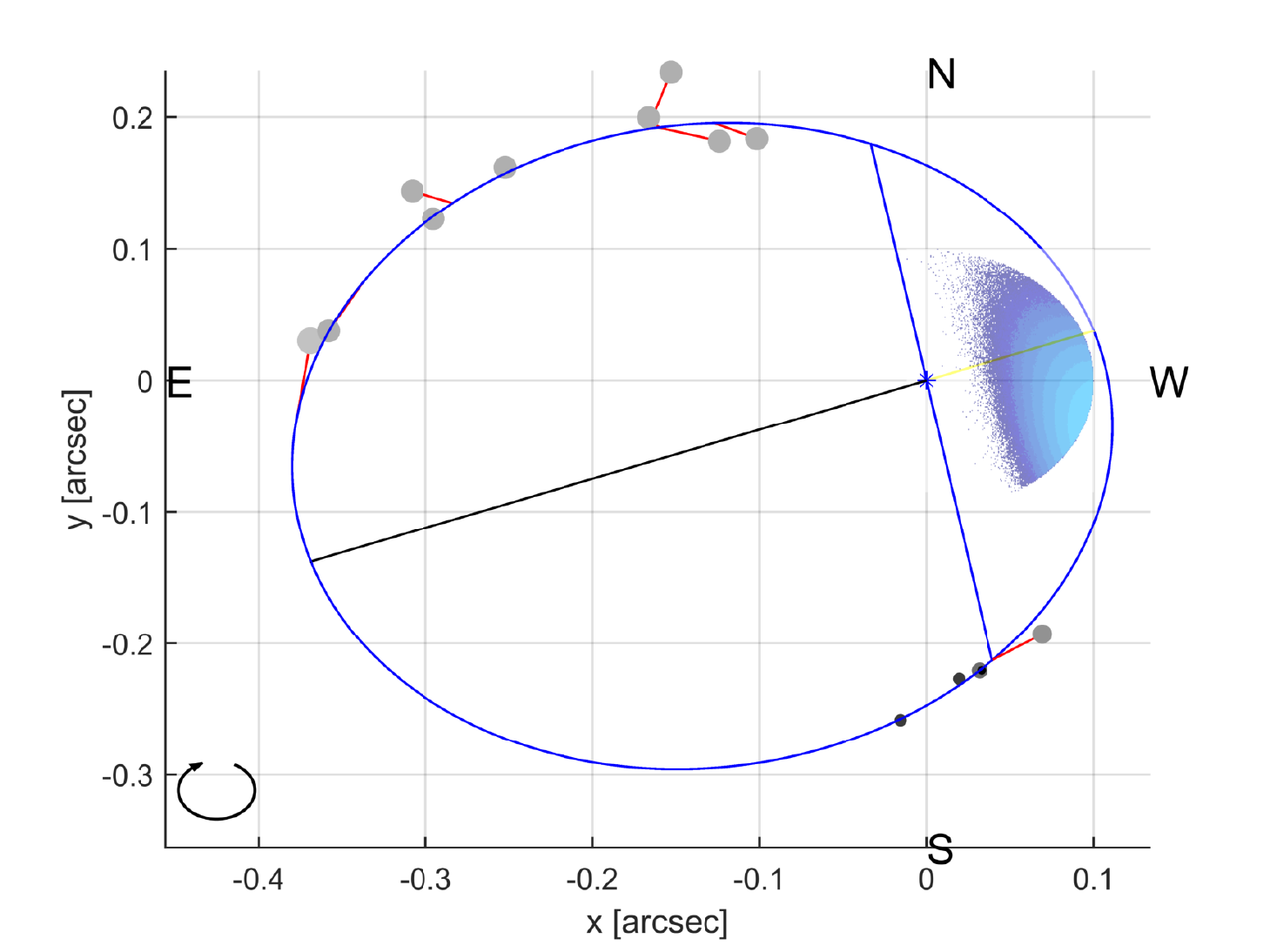}
\end{minipage}
\begin{minipage}[!ht]{0.5\linewidth}
\centering
\includegraphics[height=.25\textheight]{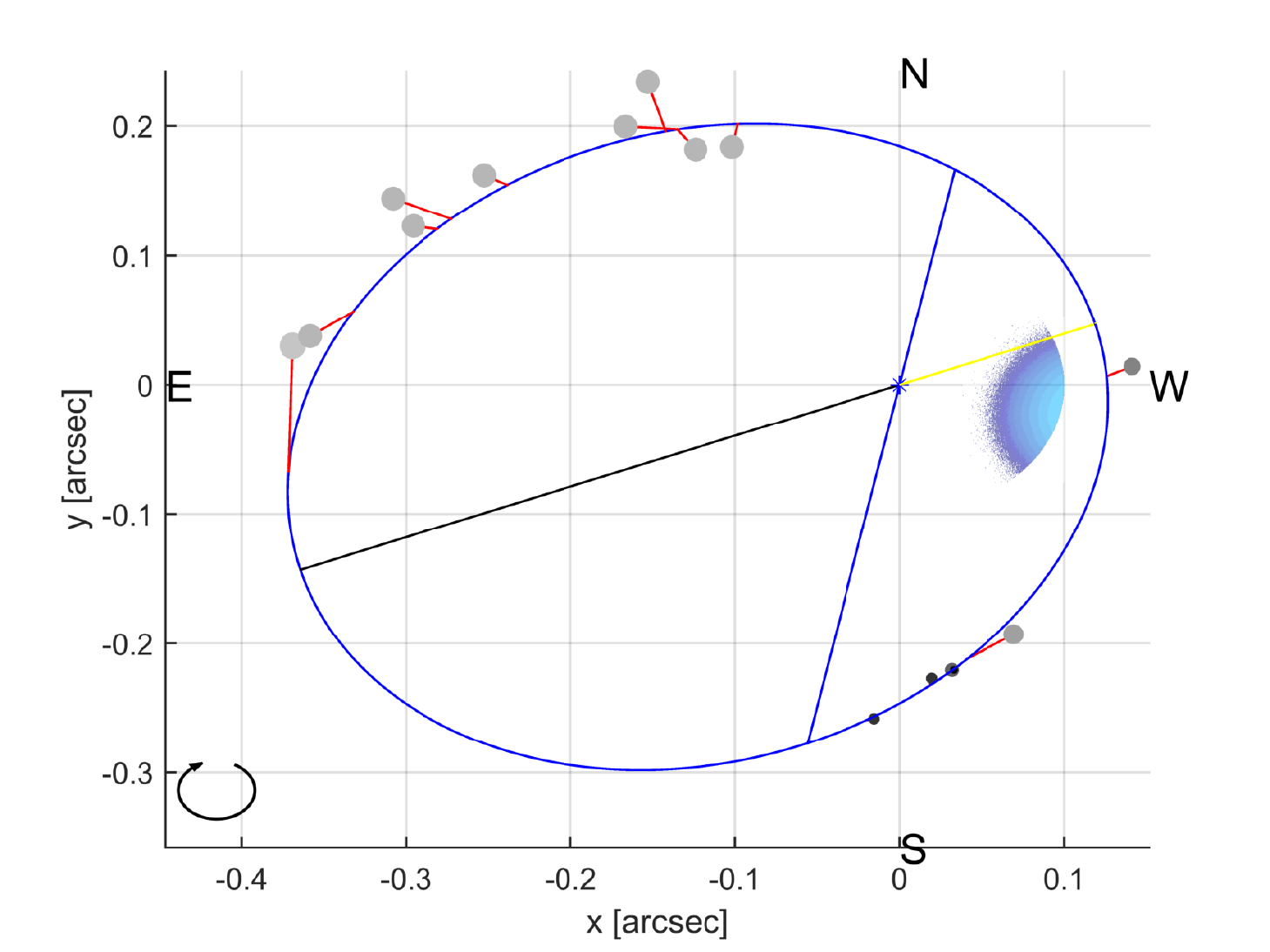}
\end{minipage} 
\caption{Comparison: visualization of the maximum likelihood orbit
  estimates for HU177$^{\ast}$ (left, with the observation near
  periastron --epoch $1989.312$-- discarded) and HU177 (right), both
  using imputation. The dots indicate the complete observations,
  larger (grey) dots imply larger uncertainties (lower weight) than
  black dots (see Table~\ref{TabHU177}). The lighter blue density
  contours indicate areas of higher probability for the imputed value
  of $\rho$ at epoch 1991.25. The line of nodes is indicated in blue
  while the line that connects periastron and apastron are yellow and
  black respectively (see color version of the figure in the
  electronic version). \label{comparisonHU177}}
\end{figure*}

From Figure~\ref{comparisonHU177} and Table~\ref{HU177_table_1} it is
observed that, although similar at first sight, orbit estimates for
HU177$^{\ast}$ and HU177 (both with multiple imputation) have
considerable, although not large, differences (see, for example,
period $P$). The estimates of angular elements $\omega$ and $\Omega$
exhibit a large numerical difference between both scenarios, but are
geometrically close due to the circular nature of angular quantities
(see Appendix~\ref{Appendix2})--in fact, both the line of nodes and
the line that connects periastron and apastron share similar
orientation and location in the two orbits. More noticeable is the
difference between the PDF of the partial measurement obtained in each
case: for HU177$^{\ast}$, it comprehends almost half the circle
defined by $\rho \in (0,\rho_\textrm{max})$ (although zones with higher
probability are concentrated towards the ``western'' border) while for
HU177, the PDF of the partial measurement is confined to a relatively
smaller area.

\begin{figure*}
\begin{minipage}[b]{0.5\linewidth}
\centering
\includegraphics[height=.83\textwidth]{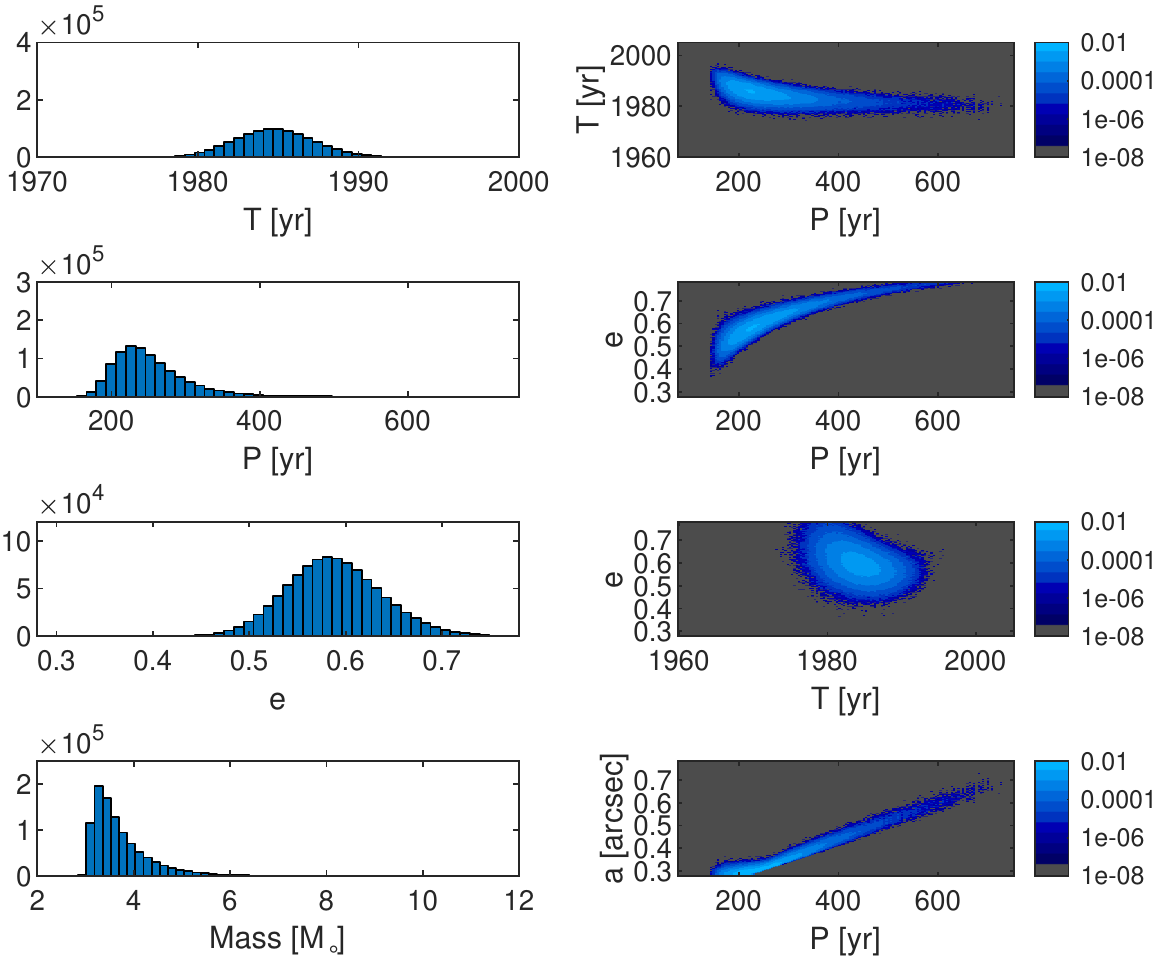}\\
\includegraphics[height=.83\textwidth]{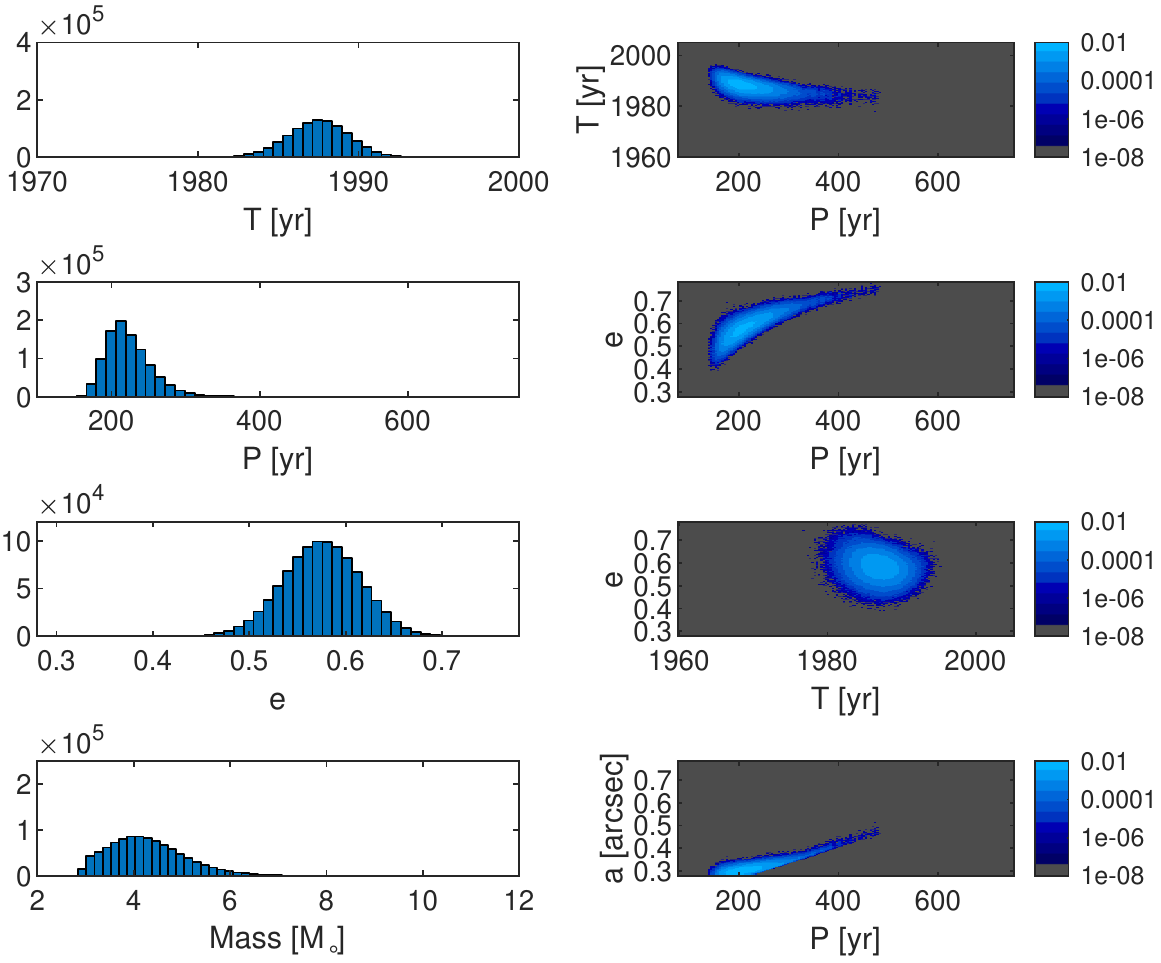}\\
\includegraphics[height=.275\textheight]{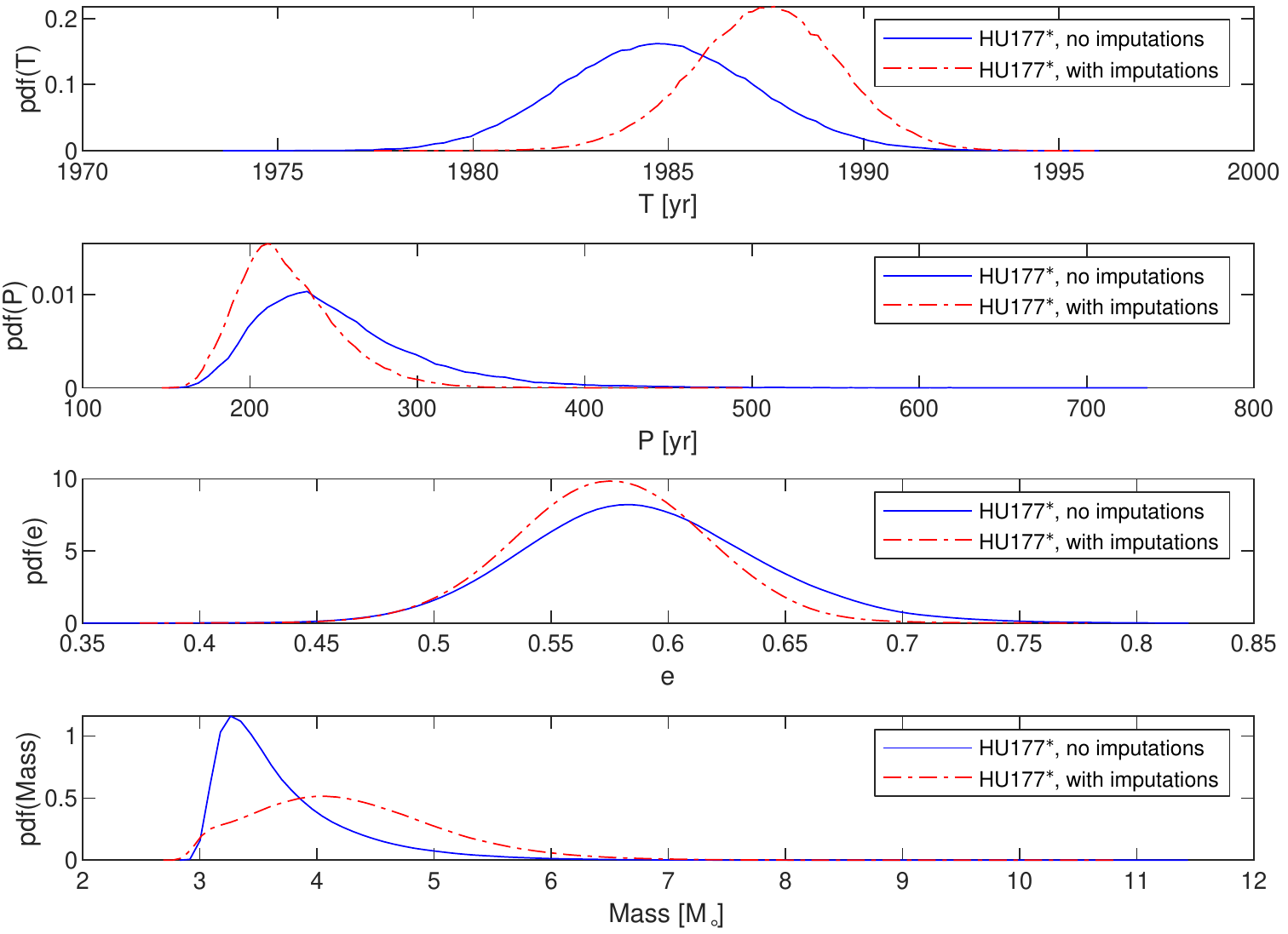}
\end{minipage}
\begin{minipage}[b]{0.5\linewidth}
\centering
\includegraphics[height=.83\textwidth]{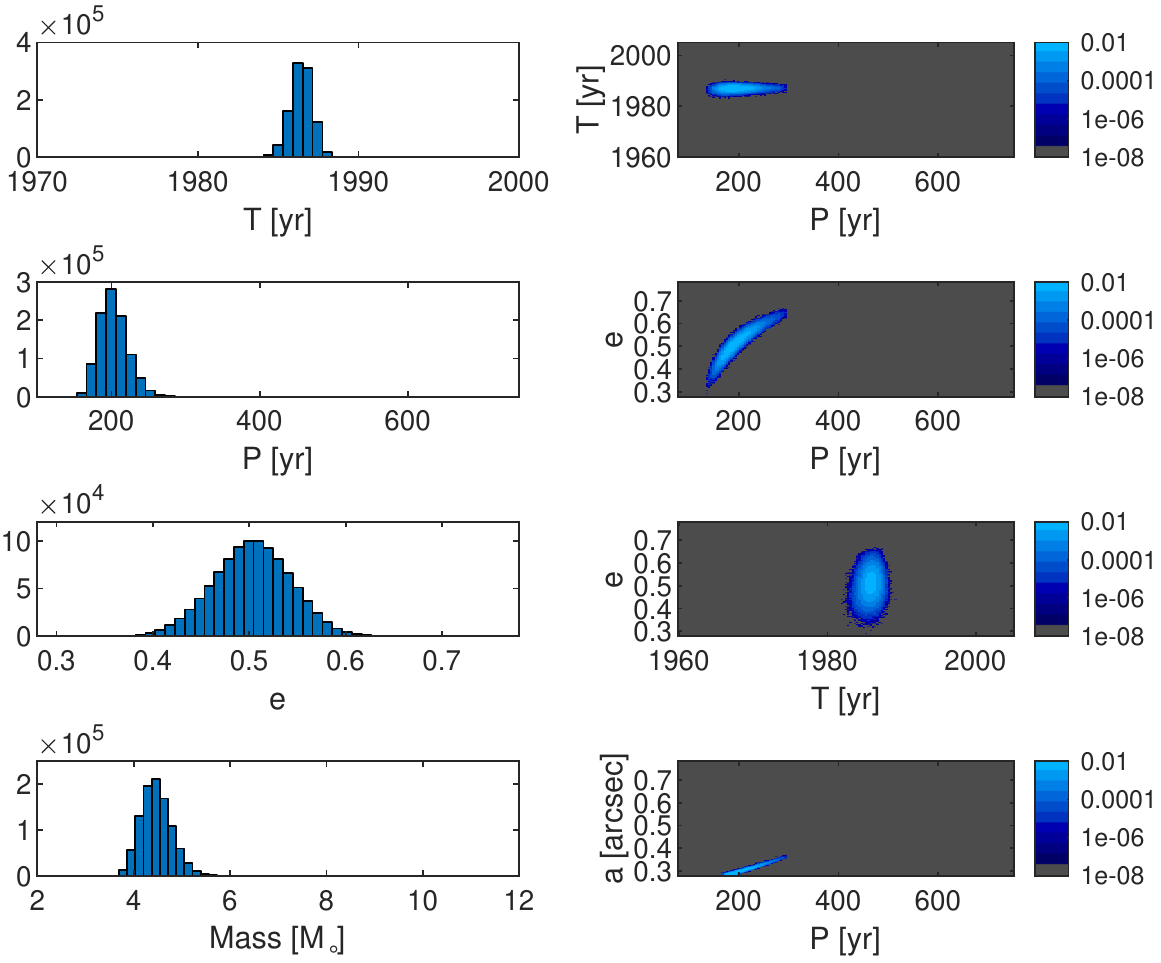}\\
\includegraphics[height=.83\textwidth]{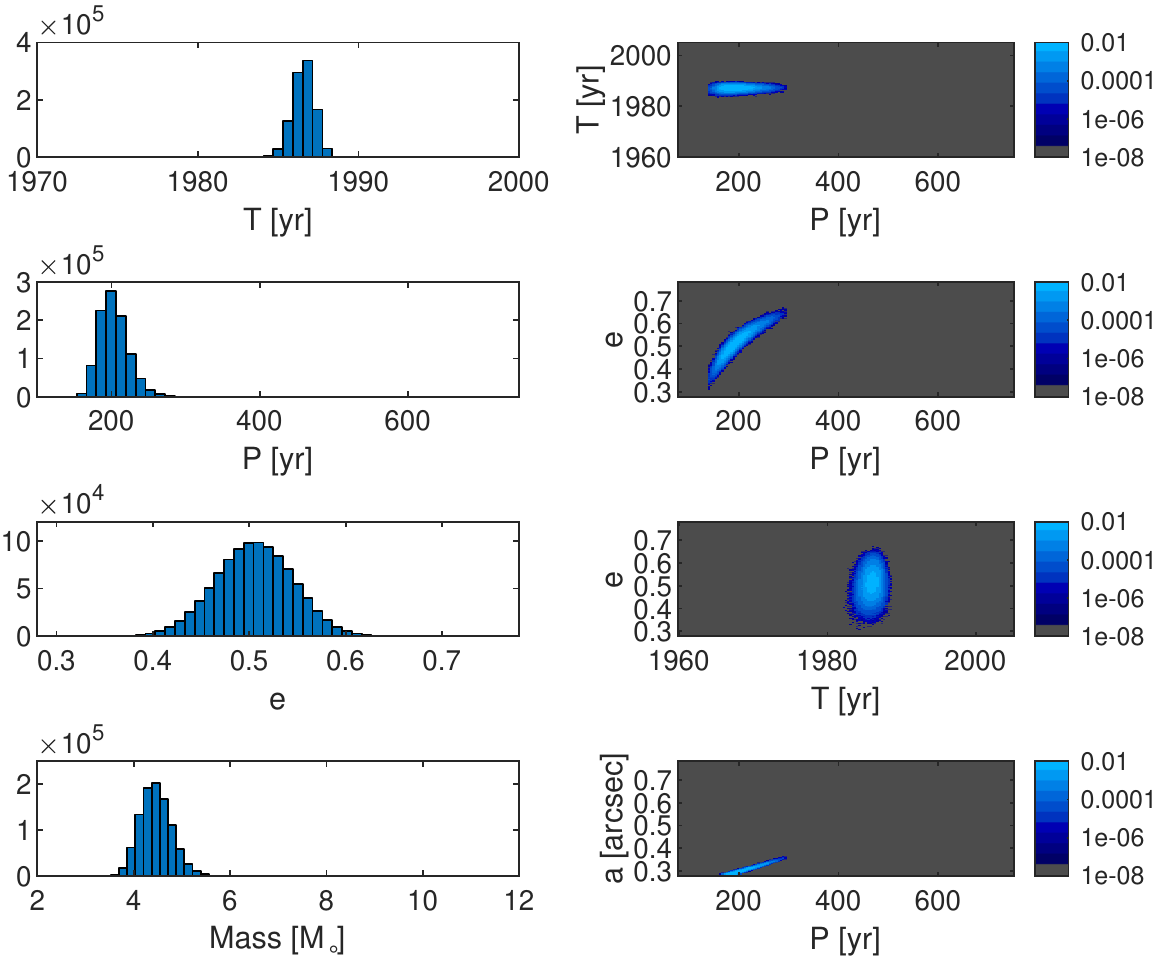}\\
\includegraphics[height=.275\textheight]{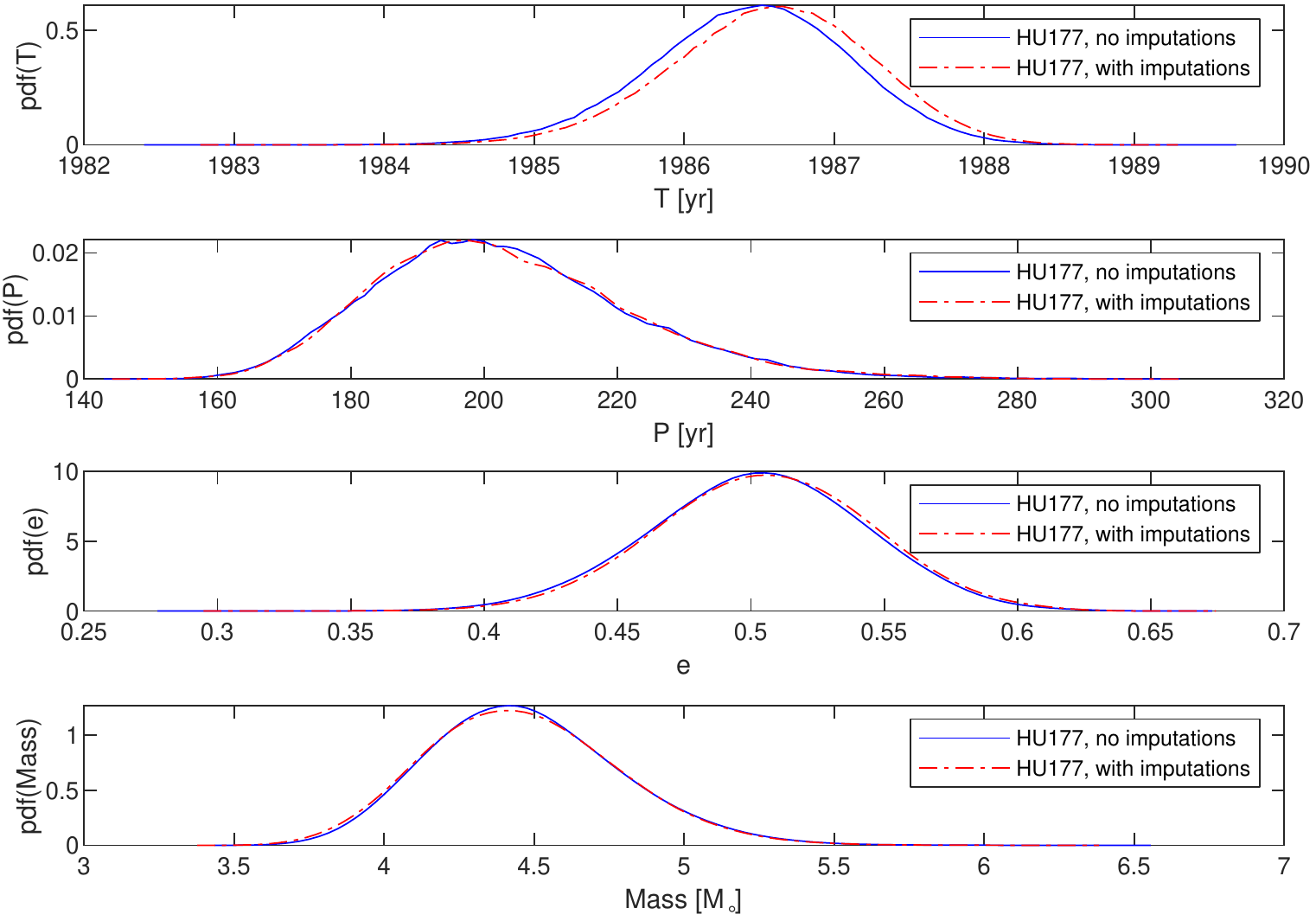}
\end{minipage}
\caption{PDFs of orbital parameters of the four cases reported in
  Table~\ref{HU177_table_2}. The left columns are for the
    HU177$^{\ast}$ data set, the right columns for the HU177 data
    set. This first four rows show the results for the case of no
    multiple imputation, while the fifth to eight rows are the results
    when using multiple imputation. The lower four panels show a
    comparison between the marginal PDFs for a selected subset of the
    orbital parameters, and the mass sum (last row). \label{HU177comparison}}
\end{figure*}

Despite the larger uncertainty about the partial observation in the
case of HU177$^{\ast}$, the incorporation of partial information has a
larger impact on the estimation in the HU177$^{\ast}$ than in the
HU177 data set: in the first setting, the interquartile range of most
parameters undergo a dramatic reduction once the multiple imputation
scheme is adopted (Table~\ref{HU177_table_2}, first two rows), whereas
in the second setting there is no statistically significant evidence
that the partial observation translates into additional knowledge of
the system\footnote{Although for some parameters the interquartile
  range has actually lower values in ``HU177 - MI'', those difference
  are within the variation range that is inherent to MCMC (due to the
  stochastic nature of the algorithm).}  (Table~\ref{HU177_table_2}, last
two rows). Since in HU177$^{\ast}$ the partial measurement is the only
source of information about that particular zone of the orbit (the
vicinity of the periastron), it contributes to reducing the
uncertainty about orbital parameters significantly. In HU177, on the
contrary, that information is redundant --it is the observation
dropped in HU177$^{\ast}$ that contributes to characterize that sector
of the orbit. Furthermore, as the imputation process in itself has
some uncertainty associated (the three sources described in
Section~\ref{MI_theory}), the multiple imputation may add uncertainty
to the estimation instead of reducing it in certain cases. That is
what apparently happens in the estimation of the total mass of
HU177$^{\ast}$, for example, where the mass sum is more constrained
when no imputations are used (Figure~\ref{HU177comparison}, fourth row
of the left panel). A question stems from the latter observation: if
the interquartile range of both $P$ and $a$ decrease when the multiple
imputation scheme is adopted, why (and how) the uncertainty of a
derived quantity such as the mass sum increases? The explanation might
be found in the fact that interquartile range is a rather na\"ive tool
to measure uncertainty, since it basically ignores the shape of the
PDFs (aspects such as heaviness of tails, symmetry, etc.). Thus,
assessing uncertainty by means of more sophisticated criteria, such as
information-theoretic indicators, appears as the next step in this
research line. Furthermore, even though the resulting PDF of the mass
sum is less concentrated, the multiple imputation estimate obtained
for that quantity ($4.17~M_\odot$) is closer to the mass estimates
obtained when no measurement is discarded ($4.42~M_\odot$); from that
point of view, the estimation using imputation improves in terms of
accuracy, even though it may not always necessarily improves in terms
of precision.

Finally, it is worth mentioning that defining more restrictive
feasible areas for missing observations (instead of mere circles and
lines) would be of great help to the estimation --the more constricted
they are, the more they resemble a well-resolved, point-like
observation. In other words, it might be beneficial, from the point of
view of parameter estimation and uncertainty characterization, to save
as much information as possible when resolving relative position
values from interferometric measurements; that means that, if complete
resolution is not feasible, an effort could be made by the observers
to confine the plausible values to, for example, an angular section
instead of a circle centered at the primary star.

\begin{deluxetable}{ccccccccc}
\tablecaption{Maximum likelihood orbit estimates for HU177 using
  incomplete information through imputation.\label{HU177_table_1}}
\tabletypesize{\normalsize}
\tablecolumns{2}
\tablewidth{0pt}
\tablehead{
\colhead{Data set} & \colhead{$P$} & \colhead{$T$} & \colhead{$e$} &
\colhead{$a$}& \colhead{$\omega$} & \colhead{$\Omega$} & \colhead{$i$}
& \colhead{Mass$_{\mbox{\tiny{T}}}$}\\
\colhead{$~$}& \colhead{$(yr)$} & \colhead{$(yr)$} &\colhead{$~$} &
\colhead{$^{\prime\prime}$} & \colhead{$(^{\circ})$} &
\colhead{$(^{\circ})$} & \colhead{$(^{\circ})$} &
\colhead{$(M_\odot)$}
}   
\startdata
HU177$^{\ast}$ & $222$ & $1987.1$ & $0.575$ & $0.296$ & $81.6$
& $10.5$ & $147.8$ & $4.02$ \\
HU177 & $205$ & $1986.6$ & $0.512$ & $0.287$ & $240.2$ &
$168.3$ & $151.4$ & $4.37$\\
\enddata
\end{deluxetable}

\begin{deluxetable}{cccccccccc}
\tablecaption{Estimation of orbital parameters for HU177 (based on
  quartiles of the PDFs) in the presence of incomplete information.\label{HU177_table_2}}
\tabletypesize{\normalsize}
\tablecolumns{4}
\tablewidth{0pt}
\tablehead{
\colhead{Data set} & \colhead{Use of multiple} & \colhead{$P$} &
\colhead{$T$} & \colhead{$e$} & \colhead{$a$} & \colhead{$\omega$} &
\colhead{$\Omega$} & \colhead{$i$} &
\colhead{Mass$_{\mbox{\tiny{T}}}$}\\ \colhead{$~$} &
\colhead{imputation} & \colhead{$(yr)$} & \colhead{$(yr)$} &
\colhead{$~$} & \colhead{$^{\prime\prime}$} & \colhead{$(^{\circ})$} &
\colhead{$(^{\circ})$} & \colhead{$(^{\circ})$} &
\colhead{$(M_\odot)$\textsuperscript{a}}
}
\startdata
HU177$^{\ast}$ & No MI\tablenotemark{b}&$243.1_{+25.0}^{-34.0}$ &$1984.7_{-1.6}^{+1.7}$ &$0.587_{-0.032}^{+0.034}$ &$0.300_{-0.013}^{+0.021}$ &$150.0_{-66.8}^{+27.5}$ &$83.9_{-72.5}^{+28.0}$ &$154.5_{-6.6}^{+6.1}$ &$3.54_{-0.24}^{+0.41}$\\
& &$59.0$ &$3.3$ &$0.066$ &$0.034$ &$94.3$ &$100.5$ &$12.7$ &$0.65$\\
HU177$^{\ast}$ & MI\tablenotemark{c} &$218.7_{+16.6}^{-21.6}$ &$1987.5_{-1.3}^{+1.2}$ &$0.575_{-0.027}^{+0.027}$ &$0.298_{-0.009}^{+0.011}$ &$84.7_{-4.2}^{+4.0}$ &$11.9_{-3.7}^{+4.7}$ &$145.4_{-5.0}^{+6.1}$ &$4.20_{-0.50}^{+0.57}$\\
& &$38.2$ &$2.5$ &$0.054$ &$0.020$ &$8.2$ &$8.4$ &$11.0$ &$1.07$\\
HU177 & No MI\tablenotemark{b} &$201.2_{+11.7}^{-13.3}$ &$1986.5_{-0.5}^{+0.4}$ &$0.503_{-0.027}^{+0.027}$ &$0.286_{-0.007}^{+0.008}$ &$237.5_{-6.5}^{+5.6}$ &$165.8_{-5.1}^{+4.4}$ &$150.9_{-2.8}^{+3.0}$ &$4.45_{-0.20}^{+0.22}$\\
& &$25.0$ &$0.9$ &$0.054$ &$0.016$ &$12.1$ &$9.6$ &$5.8$ &$0.43$\\
HU177 & MI\tablenotemark{c} &$201.2_{+11.8}^{-13.8}$ &$1986.6_{-0.5}^{+0.4}$ &$0.505_{-0.027}^{+0.027}$ &$0.286_{-0.007}^{+0.009}$ &$239.7_{-6.4}^{+5.4}$ &$167.9_{-5.0}^{+4.2}$ &$150.2_{-2.7}^{+3.0}$ &$4.44_{-0.21}^{+0.23}$\\
& &$25.5$ &$0.9$ &$0.055$ &$0.016$ &$11.8$ &$9.2$ &$5.7$ &$0.44$\\
\enddata
\tablenotetext{a}{The Hipparcos parallax for this object is $5.06 \pm
  0.97~mas$. A value of $5.06~mas$ is used for the mass calculation.}
\tablenotetext{b}{Multiple imputation strategy not applied, incomplete
  data discarded.}
\tablenotetext{c}{Incomplete data incorporated
  through multiple imputation strategy.}
\end{deluxetable}
\section{Conclusions} \label{tantan}

This work introduces a scheme for coping with partial measurements in
a Bayesian MCMC-based framework for the characterization of visual
binary star orbits. The scheme, based on filling missing or partial
measurements with a set of plausible values--the so-called multiple
imputation approach--, is tested with both synthetic and real
measurements of visual binaries. Our results in both scenarios suggest
that incorporation of partial measurements can lead to a significant
decrease in the uncertainty of the estimation of orbital parameters,
although there are also cases where partial measurements provide
negligible additional information about the orbit (that is, posterior
PDFs undergoing imperceptible changes after the incorporation of
partial data). Potentially, multiple imputation could even worsen the
quality of estimation, since the process of imputing values also
introduces some degree of uncertainty. Those concerns suggest that the
assessment of the quality of estimation by means of more sophisticated
criteria, such as the comparison of PDFs using information-theoretic
tools, should be an important part of future work in this resaerch
line.

Partial astrometric measurements intrinsically have a degree of
spatial dispersion; however, the tests performed suggest that an
interesting improvement in the estimation might arise from restricting
their geometric boundaries during the process of reducing and
reporting astrometric data, thus feeding estimation routines as the
one presented in this work with more constrained boundaries for the
unresolved observations than those typically provided.

\section*{Acknowledgements}

RMC and JSF acknowledge support from CONICYT/FONDECYT Grant
No. 1170854. MEO acknowledges support from CONICYT/FONDECYT Grant
No. 1170044. RMC, JSF, and MEO also acknowledge support from the
Advanced Center for Electrical and Electronic Engineering, Basal
Project FB0008, and CONICYT PIA ACT1405. RAM acknowledges support from
the Chilean Centro de Excelencia en Astrofisica y Tecnologias Afines
(CATA) BASAL AFB-170002.

\bibliographystyle{aasjournal}
\bibliography{refs}



\appendix

\section{On Thiele-Innes and Campbell elements} \label{appendix0}

\subsection{Least-squares estimate} \label{Appendix1}

Consider the weighted sum of individual squared errors:

$~$

\begin{equation}\label{MSEAppendix}
\sum_{k=1}^{N_{x}} \frac{1}{\sigma_x^2(k)} [X(k) - X^\textrm{model}(k)]^2 +
\sum_{k=1}^{N_y} \frac{1}{\sigma_y^2(k)} [Y(k) - Y^\textrm{model}(k)]^2
\end{equation}

Equation~\ref{proj_TI} enables us to replace $X_\textrm{model}$, $Y_\textrm{model}$
with their analytic expression for any epoch (indexed by $k$), namely:

\begin{eqnarray} \label{eqDifferenceObsComp}
X_\textrm{obs}(k) - X_\textrm{model}(k) &= X_\textrm{obs}(k) - [B\cdot x(k) + G \cdot
  y(k)]\\ \nonumber
Y_\textrm{obs}(k) - Y_\textrm{model}(k) &= Y_\textrm{obs}(k) - [A\cdot x(k) + F \cdot
  y(k)].
\end{eqnarray}

Due to the linear dependency of $X_\textrm{model}$, $Y_\textrm{model}$ with respect
to the normalized coordinates $x$, $y$, it is possible to calculate
the least-squares estimate for the unknown variables $B$, $G$, $A$,
and $F$ in a non-iterative way. Moreover, the first term of
Equation~\ref{MSEAppendix} depends only on $(B, G)$, whereas second
term depends on $(A, F)$. Therefore, the estimate for $(B, G)$ is
obtained by minimizing the first term and the estimate for $(A, F)$ by
minimizing the second one. The problem is thus reduced to a pair of
uncoupled linear equations. By calculating the derivatives of the
expression of the error with respect to each of the Thiele-Innes
constants and making the results equal to zero, one can obtain the
following formulae (for the sake of briefness, a set of auxiliary
terms is introduced first):

\begin{footnotesize}
\begin{eqnarray}
\nonumber
\alpha = \sum_i~w_i~x(i)^2 ~~~& \beta = \sum_i~w_i~y(i)^2 ~~~& \gamma =
\sum_i~w_i~x(i)~y(i)\\
~ & r_{11} = \sum_i~w_i~X_\textrm{obs}(i)~x(i) ~~~& r_{12} =
\sum_i~w_i~X_\textrm{obs}(i)~y(i) \nonumber \\
~ & r_{21} = \sum_i~w_i~Y_\textrm{obs}(i)~x(i) ~~~& r_{22} =
\sum_i~w_i~Y_\textrm{obs}(i)~y(i) \label{auxConstantsLS}
\end{eqnarray}
\end{footnotesize}

Then, the least-squares estimate for the Thiele-Innes constant is
calculated as follows:

\begin{eqnarray*}
\hat{B} = \frac{\beta \cdot r_{11} - \gamma \cdot r_{12}
}{\Delta},~~~~~~& \hat{G} = \displaystyle \frac{\alpha \cdot r_{12} - \gamma \cdot
  r_{11} }{\Delta},\\
\hat{A} = \frac{\beta \cdot r_{21} - \gamma \cdot r_{22}
}{\Delta},~~~~~~& \displaystyle \hat{F} = \frac{\alpha \cdot r_{22} - \gamma \cdot
  r_{21} }{\Delta},
\end{eqnarray*}

where $\Delta = \alpha \cdot \beta - \gamma^2$. An alternative matrix
representation is given as follows:

\begin{eqnarray}
    \begin{bmatrix}
            \hat{B}\\
            \hat{G}\\
            \hat{A}\\
            \hat{F}
        \end{bmatrix}^T = [\vec{X}_\textrm{obs} \vec{Y}_\textrm{obs}] ~\mathbf{W} \mathbf{F}^T (\mathbf{F} \mathbf{W} \mathbf{F}^T)^{-1} \text{, where } \\
        \mathbf{F} = 
    \begin{bmatrix}
            x(1) & \dots & x(N) & 0 & \dots & 0\\
            y(1) & \dots & y(N) & 0 & \dots & 0\\
            0 & \dots & 0 & x(1) & \dots & x(N)\\
            0 & \dots & 0 & y(1) & \dots & y(N)
        \end{bmatrix},
\end{eqnarray}

where $\{x(k), y(k)\}_{k = 1, \dots, N}$ are the normalized
coordinates given $P$, $T$, $e$ and epochs $\{\tau_k\}_{k = 1, \dots,
  N}$. $\mathbf{W}$ is a diagonal matrix containing the weight of each
observation (usually the reciprocal of the observational error
variance).

\subsection{Conversion from Thiele-Innes to Campbell} \label{Appendix2}

Once the estimates for Thiele-Innes constants are obtained ($\hat{B}$,
$\hat{G}$, $\hat{A}$, $\hat{F}$), it is necessary to recover the
equivalent Campbell elements representation ($a, \omega, \Omega,
i$). For $\omega$ and $\Omega$, one must solve the following
equations:

\begin{eqnarray*}
\omega + \Omega &= \arctan\left( \frac{B-F}{A+G}\right),\\
\omega - \Omega &= \arctan\left( \frac{-B-F}{A-G}\right),
\end{eqnarray*}

choosing the solution that satisfies that $\sin \left( \omega + \Omega
\right)$ has the same sign as $B - F$, and that $\sin \left( \omega -
\Omega \right)$ has the same sign as $-B-F$. If that procedure outputs
a value of $\Omega$ that does not satisfy the convention that $\Omega
\in (0, \pi)$, it must be corrected in the following way: if $\Omega <
0$, the values of $\omega$ and $\Omega$ are modified as $\omega = \pi
+ \omega$, $\Omega = \pi + \Omega$; whereas if $\Omega > \pi$, the
values of $\omega$ and $\Omega$ are modified such that $\omega =
\omega - \pi$, $\Omega = \Omega - \pi$.

For the semi-major axis $a$ and inclination $i$, the following auxiliary
variables must be calculated first:

\begin{eqnarray*}
k &= \frac{A^2 + B^2 + F^2 + G^2}{2},\\
m &= A \cdot G - B \cdot F,\\
j &= \sqrt{k^2-m^2}.
\end{eqnarray*}
Then, $a$ and $i$ are determined with the following formulae:
\begin{eqnarray*}
    a &= \sqrt{j+k}\\
    i &= \arccos \left(\frac{m}{a^2}\right)
\end{eqnarray*}

\section{Supporting pseudo-code} \label{spc}

This section outlines the fundamentals of the MCMC algorithms used in this work. Term $x$ is the variable of  interested in (e.g., a
parameter vector, the state vector of a system, etc.), $\pi(x)$ represents the target PDF and $q(x^\prime|x)$ is
a proposal distribution (a distribution to generate new samples of the
Markov chain).

\begin{figure}[!hbt]
    \caption{Metropolis-Hastings algorithm}
    \label{alg_MH}
    \centering
    \includegraphics[width=\textwidth]{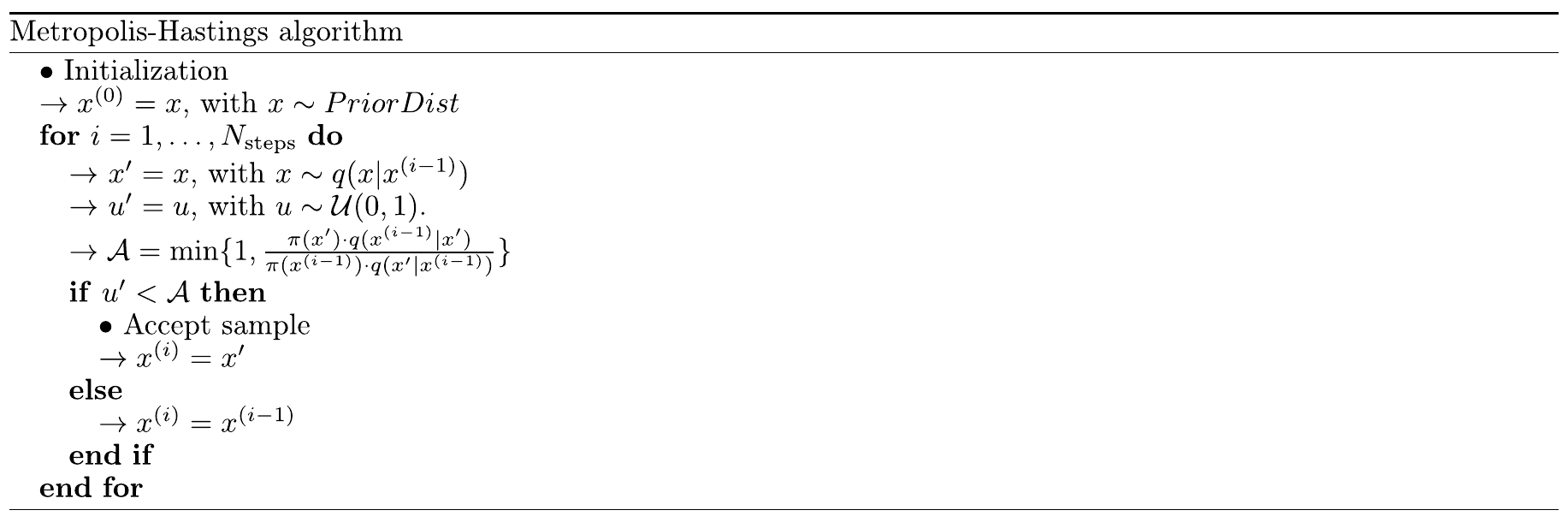}
\end{figure}

\begin{figure}[!hbt]
    \caption{Gibbs sampler \label{alg_Gibbs}}
    \centering
    \includegraphics[width=\textwidth]{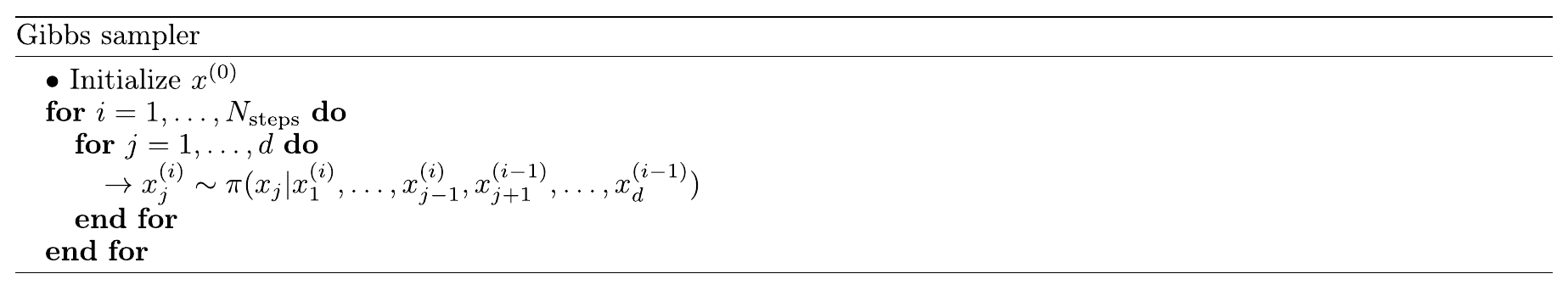}
\end{figure}

\begin{figure}[!hbt]
    \caption{Metropolis-Hastings-within-Gibbs \label{alg_MH_Gibbs}}
    \centering
    \includegraphics[width=\textwidth]{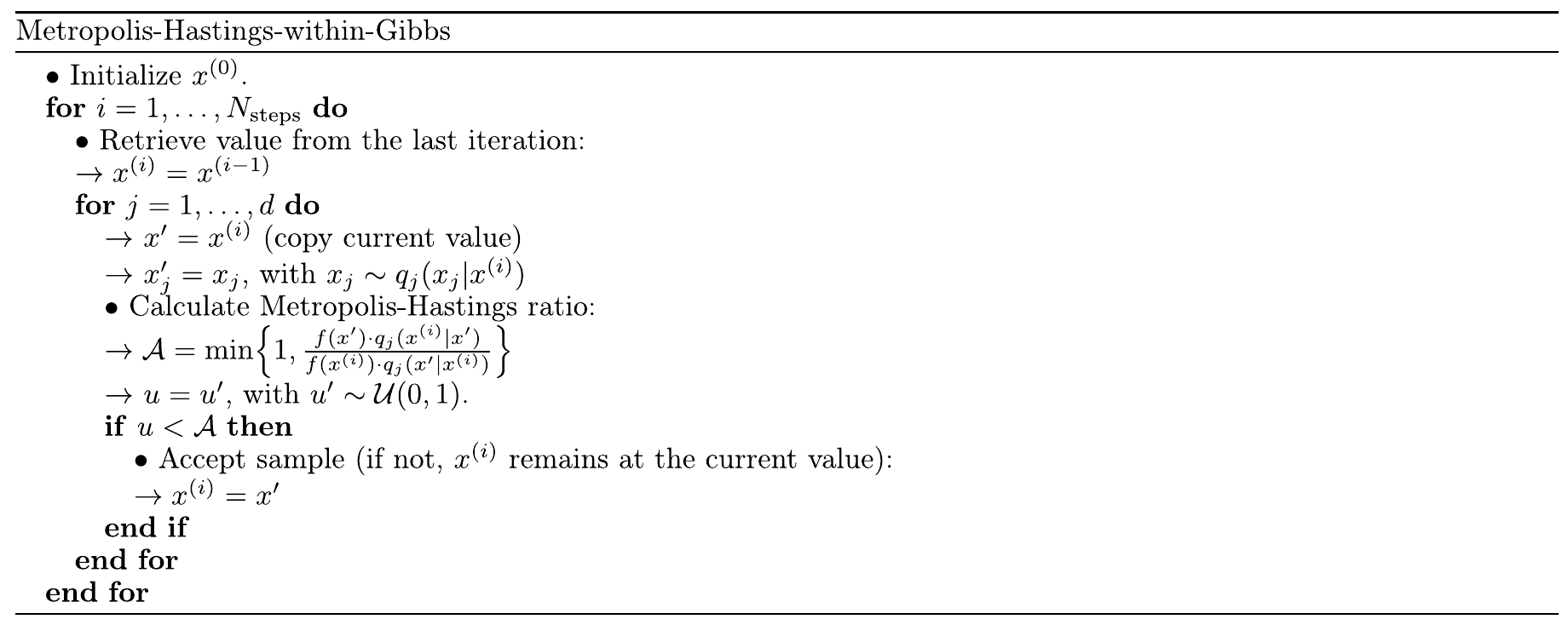}
\end{figure}

\newpage

\section{Initial state of the Markov chains and convergence diagnostics} \label{appendix3}

To assess whether the algorithm has reached a stationary regime, the convergence criterion introduced in \cite{gelman1992inference} was used. The so-called Gelman--Rubin R statistic is a tool to evaluate convergence by comparing multiple independent Markov chains. To apply the Gelman--Rubin R statistic, several chains are run, each of them starting from a different point in the parameter space. Then, after calculating quantities $B$ (inter-chain variance, Equation \ref{B_eq_gelman}) and $W$ (intra-chain variance, Equation \ref{W_eq_gelman}), $R$ statistic is computed as in Equation \ref{R_eq_gelman}.
\begin{equation}
\label{B_eq_gelman}
B = \displaystyle \frac{N_\textrm{steps}}{N_\textrm{chains}-1} \sum_{j=1}^{N_\textrm{chains}} (\hat{\theta}_j - \hat{\theta})^2.
\end{equation}
\begin{equation}
\label{W_eq_gelman}
W = \displaystyle \frac{1}{N_\textrm{chains}} \sum_{j=1}^{N_\textrm{chains}} \hat{\sigma}_j^2.
\end{equation}
\begin{equation}
\label{R_eq_gelman}
R = 1 + \displaystyle \frac{1}{N_\textrm{steps}}\cdot \biggl(\frac{B}{W} - 1\biggr).
\end{equation}
In equations \ref{B_eq_gelman}, \ref{W_eq_gelman}, \ref{R_eq_gelman}, $\theta$ denotes one of the parameters being estimated (i.e., a scalar component of the parameter vector), with $\hat{\theta}$ the average value calculated over all the chains and $\hat{\theta}_j$ the average of a single chain. The basic assumption of the R statistic is that, since theory states that all chains have the same stationary distribution, if all runs show a similar behaviour then all of them have reached the stationary regime. In the statistical literature, $R$ values less than $1.05$ are considered an indicator that the algorithm has converged.

\subsection{Synthetic data}
The algorithm to incorporate partial measurements relies on a good initialization of the parameters $P$, $T^\prime$ and $e$ (otherwise, the values of $\mathcal{Y}_\textrm{mis}$ in the imputation step may be nonsensical). To ensure that the chain starts from a feasible orbital configuration, each time we run the multiple imputation via MCMC algorithm, we draw its initial state from a preliminary chain run on the sub-set of complete measurements (rather than sampling directly from a non-informative prior). According to our results, the multiple imputation via MCMC algorithm tends to exhibit a behavior that falls between the two scenarios tested: it tends to generate parameter distributions that are are more concentrated than the ones obtained using no partial observations, but more disperse than those obtained when all complete measurements are available. On these grounds, and in order to avoid the computational burden of running several chains of the multiple imputation via MCMC algorithm, no test was performed for this algorithm.

Before running the long chain used to make the inference, we ran ten chains ($10^6$ samples long each) to assess convergence via the Gelman--Rubin R statistic. Using the Gaussian proposal distributions indicated in Section \ref{simbasedtests} and drawing initial values from $p_\textrm{prior}(T^\prime) = \mathbb{1}_{[0,1)}$, $p_\textrm{prior}(\log P) = \mathbb{1}_{[10,120)}$, $p_\textrm{prior}(e) = \mathbb{1}_{[0,0.99)}$, the test was performed on two different data sets: the sub-set of complete measurements, and the ground-truth setting where all measurements are known. The results are presented in Table~\ref{tableGelman1}, where it can be seen that the $R$ statistic is less than $1.05$ for all parameters, indicating that the chains converge (with the parameters specified in this work).

\subsection{Real data}
The test used for synthetic observations was repeated for the case of real data, with the following priors: $p_\textrm{prior}(T^\prime) = \mathbb{1}_{[0,1)}$, $p_\textrm{prior}(\log P) = \mathbb{1}_{[50,1200)}$, $p_\textrm{prior}(e) = \mathbb{1}_{[0,0.99)}$. The regular Gibbs sampler was applied on data sets HU177 and HU177$^{\ast}$ and no test for the multiple imputation via MCMC algorithm was performed. Results are shown in Table~\ref{tableGelman2}). The resultant $R$ statistic values were less than $1.05$ for all parameters, indicating that the chains converge.

\subsection{Burn-in period}
For both synthetic and real data, the burn-in period was set $10^5$ samples based on visual inspection of the test chains (i.e., those used to calculate the R statistic plus some other individual trials). In most cases fewer samples were necessary to reach the stationary distribution, but the conservative value $10^5$ was chosen thinking of a worst case scenario. These samples are removed from the raw, not the thinned chain.

\begin{deluxetable}{cccc}
\tablecaption{Gelman--Rubin R statistic for experiments with synthetic data \label{tableGelman1}} \tabletypesize{\normalsize}
\tablecolumns{2}
\tablewidth{0pt}
\tablehead{
\colhead{Data set} & \colhead{$P$} & \colhead{$T$} &
\colhead{$e$}}   
\startdata  
Complete measurements sub-set& 1.0139 & 1.0026 & 1.0013\\
Ground-truth setting& 1.0024 & 1.0024 & 1.0004\\
\enddata
\end{deluxetable}

\begin{deluxetable}{cccc}
\tablecaption{Gelman--Rubin R statistic for experiments with real data \label{tableGelman2}} \tabletypesize{\normalsize}
\tablecolumns{4}
\tablewidth{0pt}
\tablehead{
\colhead{Data set} & \colhead{$P$} & \colhead{$T$} &
\colhead{$e$}}   
\startdata        
HU177$^\ast$ & 1.0199 & 1.0192 & 1.0167\\
HU177 & 1.0332 & 1.0112 & 1.0119\\
\enddata
\end{deluxetable}

\end{document}